\documentstyle[aps,pre,epsf]{revtex}

\begin{document}
\preprint{TNT 96-ShPre2Pap-V.4}
\draft

\title{Predictability in the large:
       an extension of the concept of Lyapunov exponent}

\author{E. Aurell}
\address{Department of Mathematics, Stockholm University
          S--106 91 Stockholm, Sweden}

\author{G. Boffetta}
\address{Dipartimento di Fisica Generale, Universit\`a di Torino,
         Via Pietro Giuria 1, I-10125 Torino, Italy \\
         Istituto Nazionale Fisica della Materia, Unit\`a di Torino}

\author{A. Crisanti}
\address{Dipartimento di Fisica, Universit\`a di Roma ``La Sapienza'',
         P.le Aldo Moro 2, I-00185 Roma, Italy \\
         Istituto Nazionale Fisica della Materia, Unit\`a di Roma}

\author{G. Paladin}
\address{Dipartimento di Fisica, Universit\`a dell'Aquila,
         Via Vetoio, Coppito I-67100 L'Aquila, Italy \\
         Istituto Nazionale Fisica della Materia, Unit\`a dell'Aquila}

\author{A. Vulpiani}
\address{Dipartimento di Fisica, Universit\`a di Roma ``La Sapienza'',
         P.le Aldo Moro 2, I-00185 Roma, Italy \\
         Istituto Nazionale Fisica della Materia, Unit\`a di Roma}

\date{TNT 96-ShPrePap-V.4, 27 June 1996}

\maketitle

\begin{abstract}
We investigate the predictability problem in dynamical
systems with many degrees of freedom and a wide spectrum
of temporal scales. In particular, we study the case of 
 $3D$ turbulence at high Reynolds numbers by
introducing a finite-size Lyapunov exponent which measures the growth rate of
finite-size perturbations.
For sufficiently small perturbations this quantity coincides with 
the usual Lyapunov exponent. 
 When the perturbation is still small compared to large-scale
fluctuations, but large compared to fluctuations at the
smallest dynamically active scales, 
 the finite-size Lyapunov exponent is inversely 
proportional to the square  of the perturbation size. 
Our results are supported by numerical experiments on shell models.
We find that intermittency corrections do 
not change the scaling law of predictability.
We also discuss the relation between 
finite-size Lyapunov exponent and information entropy.
\end{abstract}


\section{Introduction}
\label{s:introduction}

The ability to predict has been the single most important qualifier of what
constitutes scientific knowledge, all since the successes of
Babylonian and Greek astronomy.
Indeed, the famous statement of Laplace that an intelligent being
with complete knowledge of the present and of the laws of Nature
will know the future for all time, assumes that
the future is completely predicated by the past,
and that perfect prediction would in
principle be possible. 
In more mathematical terms one can say that in the
physical sciences,
whether in the classical or the quantum regime,
one believes that Nature is ultimately described by differential
equations, and if one knows them and how to solve them, one knows all there
is to know about the
world\cite{Arnold}. 

Laplacian determinism is always conditioned by the fact
that in the real world initial conditions can never be known to arbitrary
accuracy.
More recent is the general appreciation of the fact
that in the presence of deterministic chaos, predictability is even more
severely limited, because small errors typically grow exponentially
in time\cite{Lorenz63}.
Most sufficiently complex systems in the world display chaos.
Therefore, most sufficiently complex systems can only be predicted
for a finite time. 
However, there may be some aspects of a system that are stable, while
others vary.
To take a familiar 
example, weather prediction is possible, typically
for about ten days on temperate latitudes, but how the wind blows on
the corner of the street is in practice unpredictable from one moment
to the next\cite{Monin}.

The words predictability and prediction are rather empty by themselves:
one has to ask for predictability of what feature against what perturbation,
in particular, against a perturbation of what size.
In complex and spatially extended systems one can typically talk 
about large-scale features and small-scale features.
The predictability of a small-scale feature against a
small-scale perturbation is typically shorter than the
predictability of a large-scale feature against a large-scale
perturbation. 
Certainly  this is not always true, and we will consider a counter example,
$2D$ turbulence, below.
But it seems to be a very common situation.
That successful large-scale prediction
essentially can mean just assuming that things do not change,
is not {\it per se} an argument against prediction in the large. Also in
weather prediction the assumption that the weather tomorrow will be
like today is a fairly good one.

In this work, a further development of results presented in
the a recent brief report \cite{ABCPV2}, we will introduce a quantity which measures 
predictability in the large, and apply it to hydrodynamic turbulence.
Before we proceed to the definitions, let us first recall some
facts about predictability in the small, i.e. 
the effects of dynamical chaos.
A system is said to be chaotic if small -- i.e. infinitesimal -- perturbations grow
exponentially in time. If the initial perturbation is of size $\delta$,
and the accepted error tolerance is $\Delta$, still small,
then a rough estimate gives that the predictability time is
\begin{equation}
  T_{\rm p}\sim \frac{1}{\lambda_{\rm max}}\ln\left(\frac{\Delta}{\delta}\right),
\label{eq:lyapunov-predict}
\end{equation}
where $\lambda_{\rm max}$ is the leading Lyapunov exponent\cite{EckmannRuelle}.

Already within the framework of infinitesimal perturbations
there are important modifications to 
(\ref{eq:lyapunov-predict})\cite{Pikovsky,Crisanti93}. 
In fact, in typical chaotic systems, 
(\ref{eq:lyapunov-predict}) is not quite true\cite{PaladinVulpiani87}.
The exponent $\lambda_{\rm max}$ is a global quantity which measures the 
{\it average} exponential 
rate of separation of nearby trajectories, and fluctuations of the local exponential 
grow should be taken into account \cite{Crisanti93}, but these effects are not what 
concerns us here.
In this paper we shall address the problem of predictability in systems with many
characteristic times, e.g. the case of fully developed turbulence where a hierarchy
of different eddy-turnover times do exist, or when the threshold $\delta$ is not small. 
In these cases the predictability time $T_{\rm p}$ is determined by the details of 
the nonlinear mechanism responsible for the growth of the error \cite{Monin,PaladinVulpiani94}. 
In particular,
$T_{\rm p}$ may have no relation with the maximum Lyapunov exponent 
 governed by the linearized 
equations for the infinitesimal error. 
In general, in this case the predictability time strongly depends
on the details of the system \cite{Monin,PaladinVulpiani87}.

According to Oseledec theorem, the leading eigenvalue of
the linearized equations of motion is $\exp(\lambda_{\rm max} t)$, except on
a set of points of measure zero.
The sub-leading eigenvalues have the form $e^{\lambda_i t}$, and, taken together,
the leading and sub-leading Lyapunov exponents measure the growth
rate of $d$-dimensional volumes spanned by $d$ infinitesimal vectors,
where $d$ can range from one to the dimensionality of the space where the
motion takes place.

In dynamical systems, in addition to Lyapunov exponents, an important dynamical
characterization is given by the Kolmogorov-Sinai entropy,
which measures the bandwidth necessary to observe a system over
time, so that it could later be faithfully reproduced from the
observations\cite{EckmannRuelle}.
Arguing heuristically, new observations are necessary if an error
grows in time, and the necessary rate of accumulation of information
is the growth rate of the error, and therefore one expects 
Pesin theorem,
\begin{equation}
      h_{\rm KS} = \sum_{\lambda_i>0} \lambda_i,
\label{eq:pesin-theorem}
\end{equation}
to hold true for a large class of systems.

It was realized by Shannon
that with finite error tolerance, the relevant
quantity is the bandwidth necessary to observe a system such that it
could later by reproduced within this error, not to arbitrarily
high accuracy.
This quantity was called
by ``source entropy with respect to a fidelity
criterion'' by Shannon\cite{Shannon49}, and 
``$\epsilon$-entropy''
by Kolmogorov\cite{Kolmogorov56},
who found analytic formulae, valid for Gaussian variables
and Gaussian stationary processes.
Recently the concept of $\epsilon$-entropy was taken up
by Gaspard and Wang, who computed it from experimental data
in thermal turbulence\cite{GaspardWang92}, and for a very large
variety of model problems in stochastic processes and statistical
physics\cite{GaspardWang93}. We especially recommend their lucid
and remarkably complete review\cite{GaspardWang93}.

However, the $\epsilon$-entropy does not say all one wants to know
about predictability in the large. Just as the Lyapunov exponent
is often more relevant, and more easily computable, than the
Kolmogorov-Sinai entropy, so the predictability time with respect
to a finite perturbation should be determined by a quantity analogous
to the Lyapunov exponent, and not by the $\epsilon$-entropy.

The natural starting point in looking for such a quantity is the
time it takes for a perturbation to grow from an initial size
$\delta$ to a tolerance $\Delta$. We call this the $(\delta,\Delta)$
predictability time and denote it by $T(\delta,\Delta)$.
Generally speaking, the predictability time will fluctuate.
The natural definition of the finite-size Lyapunov exponent is therefore
an average of some function of the predictability time, such that if
both $\delta$ and $\Delta$ are in the infinitesimal range, we will recover
the usual Lyapunov exponent, and an obvious choice is then
\begin{equation}
     \lambda(\delta,\Delta) =
         \left\langle\frac{1}{T(\delta, \Delta)}\right\rangle
                    \ln\left(\frac{\Delta}{\delta}\right).
\label{eq:predict}
\end{equation}
In appendix \ref{app:A} we discuss other possible definitions for the finite-size 
Lyapunov exponent and the relation with the $\epsilon$-entropy.

In contrast to infinitesimal perturbations, for finite perturbations
the threshold $\Delta$ is typically not to be taken much larger
than the perturbation $\delta$. 
What is interesting, and what makes finite-size Lyapunov exponents different
from Lyapunov exponents for infinitesimal perturbations, is the dependence
on~$\delta$.

This paper is organized as follows: in section \ref{s:multifractal}
we recall the multifractal approach to turbulence, and Lorenz approach
to the predictability problem within the Kolmogorov theory.
We show that there are no multifractal corrections to the results
of Lorenz, but that the scaling range for the finite-size Lyapunov exponents
is shorter.
In section \ref{s:shell-model} we describe numerical experiments on predictability
in shell models for $3D$ and $2D$ fully developed turbulence.
In section \ref{s:EDQNM} we present the results from the eddy damped quasi normal 
Markovian (EDQNM) approximation for the shell model, and compare them with the results
of section \ref{s:shell-model}. In section \ref{s:conclusion} we summarize our
results and present
conclusions. In appendix~\ref{app:A} we discuss alternative ways of defining
the finite-size Lyapunov exponent. Appendix~\ref{s:kolmogorov} contains
a derivation of the Kolmogorov results of $\epsilon$-entropy for Gaussian
processes and Gaussian random fields.
In appendix~\ref{s:spacetime} we apply the results of appendix~\ref{s:kolmogorov} 
to space-time Gaussian fields with spectra as in Kolmogorov 1941 theory of
$3D$ turbulence,
and to fictitious $0$-dimensional fields that describe shell models.
We show that the finite-size Lyapunov exponent is the the same for
any dimensionality, but the $\epsilon$-entropy  largely depends on
the dimensionality-dependent density of degrees of freedom.
The results of sections \ref{s:multifractal} and \ref{s:shell-model}
were also presented in less generality in \cite{ABCPV2}.

\section{Multifractals and multiscaling}
\label{s:multifractal}

Our understanding of high Reynolds-number turbulence
is still mainly based on the fundamental contribution
of Kolmogorov in 1941\cite{Kolmogorov41}.
We will here just discuss on a phenomenological level
the predictions of the Kolmogorov theory and of
its multifractal generalizations
\cite{ParisiFrisch,BPPV,Frischbook}.

Turbulence is a statistically stationary state
of matter on macroscopic scales maintained by external forces.
One considers only effects that are
captured in a hydrodynamic level of description,
that is the time evolution is supposed to be
completely described by the macroscopic
Navier-Stokes 
equations\cite{Landau-Lifshitz-Fluid-Mech}
\begin{eqnarray}
\partial_t \bbox{v} + (\bbox{v}\cdot\nabla) \bbox{v} &=& 
                                        -\nabla P + \nu \nabla^2 \bbox{v}, 
\label{eq:Navier-Stokesa} \\
                                \nabla\cdot \bbox{v} &=& 0.
\label{eq:Navier-Stokesb}
\end{eqnarray}
 From the typical length scale
$L$, the typical fluctuations of velocity on
that scale $V$, and the viscosity $\nu$, we can form the
Reynolds number,
\begin{equation}
      \hbox{Re} = {{L V}\over {\nu}},
\label{eq:Reynolds}
\end{equation}
which characterizes the flow. 

The multifractal model \cite{ParisiFrisch,BPPV,Frischbook,PaladinVulpiani87} 
consists in assuming that at
scales much less than $L$, however sufficiently 
large that the action of viscosity is weak, the velocity differences 
assume a scaling form
\begin{equation}
|\bbox{u}(\bbox{x}+\bbox{l})-\bbox{u}(\bbox{x})| = 
                    u_l \sim V \left(\frac{l}{L}\right)^h.
\label{eq:scaling-ansatz}
\end{equation}
Different values of $h$ are assumed to occur according to a 
probability distribution, which also takes a scaling form
\begin{equation}
   \hbox{Prob}\left\{\frac{u_l}{V}\in \left[\left( \frac{l}{L}\right)^h,
                     \left(\frac{l}{L}\right)^{h+dh}\right] \right\}    \sim 
                     \left(\frac{l}{L}\right)^{3-D(h)} dh.
\label{eq:probability-ansatz}
\end{equation}
The function $D(h)$ is the fractal 
 dimension of the subset with scaling exponent $h$.
The moments of the velocity differences on length scale 
$l$ can be computed as
\begin{equation}
\langle|\bbox{u}(\bbox{x}+\bbox{l})-\bbox{u}(\bbox{x})|^q \rangle \sim
                V^q \int\left(\frac{l}{L}\right)^{qh} \left(\frac{l}{L}\right)^{3-D(h)} dh,
\label{eq:moment-q}
\end{equation}
and, for small-$l$ (i.e. in the inertial range), 
the integral in (\ref{eq:moment-q})
can be evaluated by saddle point method:
\begin{eqnarray}
 \langle|\bbox{u}(\bbox{x}+\bbox{l})&-&\bbox{u}(\bbox{x})|^q \rangle \sim 
             V^q  \left(\frac{l}{L}\right)^{\zeta_q},\\
\zeta_q &=& \min_h \bigl[qh +3-D(h)\bigr].
\label{eq:zeta-q}
\end{eqnarray}

The model is physically reasonable for a large set of possible
choices of the function $D(h)$, but not entirely arbitrary.
By normalization, the value of $D(h)$ must always be less
or equal than three, and the maximum must be obtained for
some $h$. That is
\begin{eqnarray}
       3-D(h) \ge 0. 
\label{eq:zeta-0-inequality}
\end{eqnarray}
The function $D(h)$ can have support only at positive $h$,
because a negative value of $h$ implies that velocity fluctuations
in a local inertial frame of size $l$ increase without
limit as $l$ tends to zero. The Navier-Stokes equations
are derived under the assumption that all velocities are
much smaller than the velocity of sound, and this condition
would then no longer hold \cite{Frischbook}.
Furthermore, an exact result of Kolmogorov assures that 
$\zeta_3=1$, so that 
\begin{eqnarray}
    3h +3-D(h) \ge 1, 
\label{eq:zeta-3-inequality}
\end{eqnarray}
where equality is obtained for at least one value of $h$\cite{Frischbook}. 
The inequality (\ref{eq:zeta-3-inequality}) is the analogous for turbulence of 
the inequality $f(\alpha)\leq\alpha$ for multifractal measures \cite{PaladinVulpiani87}.

In terms of the multifractal model, the Kolmogorov theory is
formulated by supposing that the function $D(h)$ has support
only at a single point. From (\ref{eq:zeta-0-inequality}) and
(\ref{eq:zeta-3-inequality}) it then follows that this point
must be $h=1/3$, and that $D(1/3) = 3$.
It further follows the Kolmogorov  law
\begin{equation}
  \langle|\bbox{u}(\bbox{x}+\bbox{l})-\bbox{u}(\bbox{x})|^p \rangle
                \sim V^p\left(\frac{l}{L}\right)^{p/3}.
\label{eq:Kolmogorov-41}
\end{equation}
Energy dissipation per unit mass and time, $\epsilon$,
has dimension $V^3/L$.
We could therefore also write the right-hand side
in  (\ref{eq:Kolmogorov-41}) in the more
familiar form~$(\epsilon l)^{p/3}$

 From (\ref{eq:Kolmogorov-41}) it follows
by balancing in (\ref{eq:Navier-Stokesa}), that viscous forces become
comparable with inertial forces 
at the Kolmogorov scale $\eta$ which marks the lower end of the inertial
range:
\begin{equation}
  \eta = L\,  \hbox{Re}^{-3/4}.
\label{eq:Kolmogorov-scale}
\end{equation}
If there exists more than one value of $h$ then each $h$ selects a different damping scale
$\eta(h)$. By using (\ref{eq:scaling-ansatz}) and balancing, one gets 
\cite{palvul} 
\begin{equation}
     \eta(h) = L\,  \hbox{Re}^{-1/(1+h)}.
\label{eq:Frisch-Vergassola-scale}
\end{equation}

Lorenz investigated the predictability problem within the
framework of the Kolmogorov theory\cite{Lorenz69}. Let us assume that we
have a disturbance on scale $l$, and that it grows
at a characteristic rate given by the turn-over time 
at this scale:
\begin{equation}
  \tau(l) = {l\over{\sqrt{\langle u_l^2\rangle}}} 
          \sim \frac{L}{V}\left(\frac{l}{L}\right)^{2/3}.
\label{eq:turn-over-time}
\end{equation}
We can turn around (\ref{eq:turn-over-time}) and say that after
a time $t$ a disturbance will have grown large on all scales
smaller than 
\begin{equation}
     l(t) \sim L  \left(\frac{Vt}{L}\right)^{3/2}.
\label{eq:decorrelated-scale}
\end{equation} 
The size of the disturbance will then be $\langle u_{l(t)}^2\rangle^{1/2}$,
since all the smaller scales contribute relatively little.
If we call the difference between two fields $\delta$, we can
rewrite (\ref{eq:turn-over-time}) as the predictability time
with respect to a perturbation of size $\delta$:
\begin{equation}
 \tau(\delta) \sim \frac{L}{V}\left(\frac{\delta}{V}\right)^{2}.
\label{eq:Lorenz-predictability-time}
\end{equation}
In other words, the predictability time of a perturbation of
size $\delta$ grows as $\delta^2$ in Lorenz scenario.
The finite-size Lyapunov exponent thus decreases 
with the error threshold as $\delta^{-2}$.
Finally, we can insert $\delta$ in (\ref{eq:decorrelated-scale}),
and find how the error grows with time:
\begin{equation}
   \delta(t) \sim V \sqrt{\frac{Vt}{L}}.
\label{eq:error-growth-rate}
\end{equation} 
The upshot of these simple estimates is that finite error growth
and predictability in high Reynolds number turbulence are
characterized by {\it algebraic} laws, very different 
form the {\it exponential} laws characteristic of infinitesimal perturbations in 
chaotic dynamical systems.

We now turn to possible consequences of a spectrum of $h$'s
to the predictability problem, in direct analogy with 
the Lorenz theory.
We start by assuming the inverse of
(\ref{eq:probability-ansatz}) that is
\begin{equation}
    \hbox{Prob}\left\{\frac{\delta}{V}\in \left[\left(\frac{l}{L}\right)^h,
                 \left(\frac{l}{L}\right)^{h+dh}\right] \right\} 
               \sim \left(\frac{\delta}{V}\right)^{\frac{3-D(h)}{h}}dh,
\label{eq:probability-ansatz-inverse}
\end{equation}
where we have identified $u_l$ with $\delta$, and 
used (\ref{eq:scaling-ansatz}) to relate $l$ and $\delta$.
The length $l$ is now assumed a fluctuating quantity, and has the interpretation
of a scale such that two fields are uncorrelated on all smaller scales,
given that the distance of the two configurations is $\delta$.

The finite-size Lyapunov exponent is, according to (\ref{eq:predict}),
proportional to
the expectation value of the inverse predictability time:
\begin{equation}
 \left\langle\frac{1}{T(\delta)}\right\rangle = 
 \left\langle\frac{u_l}{l} \right\rangle \sim
     \frac{V}{L}\,
     \int \left(\frac{\delta}{V}\right)^{1-\frac{1}{h}}
     \left(\frac{\delta}{V}\right)^{\frac{3-D(h)}{h}} dh.
\label{eq:multifractal-predict}
\end{equation}
In the small error limit we thus expect the finite-size Lyapunov
exponent to scale as a power of the error size:
\begin{eqnarray}
 \left\langle\frac{1}{T(\delta)}\right\rangle &\sim&  \frac{V}{L}
           \left(\frac{\delta}{V}\right)^{\chi},\\
      \chi &=& \min_h \bigl[1+\frac{2-D(h)}{h}\bigr].
\label{eq:chi}
\end{eqnarray}
The exponent $\chi$ is always equal to the Lorenz value
of $-2$. This follows easily from
(\ref{eq:zeta-3-inequality}), which can be rewritten in the form
\begin{equation}
  1 +\frac{2-D(h)}{h} \ge -2\qquad\hbox{for all $h$},
\label{eq:chi-inequality}
\end{equation}
where the equality holds for the exponent
$h_3$ which dominates the third order structure function.

As far as we know this result is new.
One could therefore conclude that
the exponent $\chi$ for the scaling of the finite-size Lyapunov exponent
with error threshold is a new invariant of the
multifractal approach to turbulence, and that the law
$\langle 1/T(\delta)\rangle\sim\delta^{-2}$
can be easily observed in numerical experiments.
This is not quite simple, due to the influence of the fluctuating
cut-off (\ref{eq:Frisch-Vergassola-scale}).
The smallest fluctuation in a field scaling with exponent $\tilde{h}$
is
\begin{equation}
  \delta(\tilde{h}) = V\, \hbox{Re}^{-\tilde{h}/(1+\tilde{h})},
\label{eq:fluctuation-scale}
\end{equation}
which inversely determines the smallest value of $h$ contributing
to a fluctuation of size $\delta$.
A modified version of (\ref{eq:multifractal-predict})
therefore reads
\begin{equation}
 \left\langle\frac{1}{T(\delta)}\right\rangle =
    \left\langle \frac{u_l}{l} \right\rangle \sim
    \frac{V}{L}
   \int_{\tilde{h}}^{h_{\hbox{\scriptsize max}}}
   \left(\frac{\delta}{V}\right)^{1-1/h}
   \left(\frac{\delta}{V}\right)^{(3-D(h))/h} dh.
\label{eq:multifractal-predict-intermediate}
\end{equation}
The integral is dominated by
$h_3$  as long as
$\delta$ is much larger
than $\delta^*\sim V \hbox{Re}^{-h_3/(1+h_3)}$.
For smaller $\delta$  values, the integral is dominated
by the lower end-point
in (\ref{eq:multifractal-predict-intermediate}),
which leads to an intermediate
dissipative range, in the sense of Frisch and
Vergassola\cite{FrischVergassola}. 
As a consequence, we have
\begin{equation}
\label{eq:multi-verg} 
\left\langle \frac{1}{T(\delta)}\right\rangle \sim \left\{\begin{array}{lcl}
           \lambda_{\rm max} &\hbox{for}& \delta <  \delta(h_{\rm max}), \\
           \delta^{\,\chi(\delta)} &\hbox{for}& \delta(h_{\rm max}) < \delta < \delta^*\\
           \delta^{-2} &\hbox{for}& \delta > \delta^*,  \\
        \end{array}\right.
\end{equation}
with
\begin{equation}
\label{eq:chi1}
     \chi(\delta) = 1 + \frac{2 - D(\tilde{h})}{\tilde{h}}, 
\end{equation}
where $\tilde{h}$ and $\delta$ are related via (\ref{eq:fluctuation-scale}). From 
(\ref{eq:zeta-3-inequality}) follows that $h_3$ is less or equal
to $1/3$ in the multifractal approach, and therefore
the bottom of the scaling range of the finite-size
Lyapunov exponent is larger than the corresponding error size
in the Kolmogorov theory.
The scaling range for the finite-size Lyapunov exponent thus is 
 generally  shorter in a multifractal model.

\section{Error growth in shell models}
\label{s:shell-model}

Simplified dynamical  models of fluid turbulence with relatively
few degrees of freedom, collectively referred to as {\it shell models},
have been studied since the seventies.
 These models by construction typically include a
Richardson cascade of energy from large to small scales.
Some of the models are dynamically stable, with
a fixed point which reproduces the Kolmogorov 
 law for the energy spectrum, $E(k)\sim k^{-5/3}$.
An historical overview, with references to much of the early
work, can be found in a recent
monograph \cite{turbulence-book}.

More interestingly, other models are dynamically unstable,
with chaotic motion taking place on a strange attractor where the Kolmogorov
$5/3$ law holds to good accuracy, but not exactly.
One of the  simplest example
 is  the family of models introduced
by Gledzer\cite{Gledzer}, and Yamada and Ohkitani \cite{YamadaOhkitani87},
now commonly called the GOY models\cite{Pisarenko93}.
They have 
recently been the subject of several investigations
\cite{YamadaOhkitani88a,JensenPaladin91,Crisanti93,BKLMW,BLLP}.
An in-depth description of this work can also be found in
\cite{turbulence-book}.

The  GOY models are defined as follows:
Fourier space is divided into $n=1,\ldots,N$ shells, labeled
by the wave-vector 
modulus
$k_n = k_0\,2^n$,  where $k_0$ is a constant.
The velocity difference over a length
scale $l_n \sim k_n^{-1}$ are represented, each by
one complex variable $u_n$, which obey the following system of coupled
ordinary differential equations:
\begin{eqnarray}
 \frac{d}{dt} u_n &=& - \nu k_n^2 + i g_n + f\,\delta_{n,4}
\label{eq:sh1} \\
  g_n &=& a\, k_n\,     u_{n+1}^*\, u_{n+2}^* +
          b\, k_{n-1}\, u_{n-1}^*\, u_{n+1}^* +
          c\, k_{n-2}\, u_{n-2}^*\, u_{n-1}^*
\label{eq:sh3}
\end{eqnarray}
where $f$ is the strength of the external force, acting on large scales, and 
$\nu$ the viscosity.
For any values of the three coefficients $a$, $b$ and $c$, phase space
volume in preserved in the force-free inviscid limit.

The restricted number of degrees of freedom is both the main advantage
and disadvantage of shell models. It is an advantage, because it allows
simulations at much lower viscosity and for much longer time than
in the full Navier-Stokes equations. But it is also
a severe departure.
All spatial structure of the field is ignored. 

One of the coefficients in (\ref{eq:sh3}), say $a$, can  be scaled to one
and the condition of
energy conservation fixes one more, such that, in terms of
one parameter $\epsilon$,
$b$ is equal to $-\epsilon$ and
$c$ to $-(1-\epsilon)$.
With $\epsilon$ greater than one the GOY equations conserve
one more positive definite quantity besides energy, i.e. an
analogous situation to $2D$ hydrodynamics.
The dynamical behavior of the GOY models in this range is
rather far removed from $2D$ turbulence as was shown in recent papers\cite{Aurell94,DitlevsenMorgensen}.
We include below a study of predictability in such shell models,
but just as a simplified model to demonstrate one possible scaling behavior
of the finite-size Lyapunov exponent.
 
With parameter $\epsilon$ in the range between and zero and one,
the GOY models also preserve another invariant, but which is
not positive definite.
In the following we will look at the GOY model with
the standard choice of $\epsilon$ equal to $1/2$, i.e.
\begin{equation}
  a = 1, \qquad b = -\frac{1}{2}, \qquad c = -\frac{1}{2}.
\label{eq:GOY_standard}
\end{equation}
The second invariant then has physical
dimension of helicity\cite{BKLMW}.
Presumably that is the reason why this particular model has turned out
to be so close to numerical and experimental data on
Navier-Stokes turbulence in $3D$ 
\cite{BiferaleKerr,BenziBiferaleTrovatore}.

Before we turn to the numerical experiments, let us summarize
some salient features of the system defined by 
(\ref{eq:sh1})-(\ref{eq:GOY_standard}).
Energy is pumped into the system by the force, which acts only
on shells with low values of $n$, and is removed at high shells
by viscosity. From $k_0$, $\nu$ and the typical fluctuations of velocity
on large scales, $V$, we can form a Reynolds number
$\hbox{Re} = V/k_0\nu$.
We consider the situation where $\hbox{Re}$ is large,
such that there is a wide range in $n$ where the external and viscous
forces are both negligible compared to the inertial forces.
In this inertial range,
we have $\langle |u_n|^2\rangle\sim  k_n^{-\zeta_2}$,
where the exponent $\zeta_2$ is close to $2/3$\cite{JensenPaladin91}.

An estimate of the smallest excited scale is in analogy with the 
Kolmogorov scale $k_{n^*}\sim k_0 \hbox{Re}^{3/4}$.
The slowest dynamical scale is the time-scale of the 
the shells containing most energy, about $1/k_0V$, and
the fastest is
$\tau_{n^*}^{-1} \sim k_{n^*}\langle |u_{n^*}|^2\rangle ^{1/2}$,
or, about $k_0V\hbox{Re}^{1/2}$. From the fastest time-scale follows by dimensional 
analysis in the
Kolmogorov theory that the leading Lyapunov exponent should
grow with Reynolds number
as $\hbox{Re}^{1/2}$, a prediction due to Ruelle\cite{Ruelle79}.
In the multifractal picture there are corrections to this estimate
and that the leading Lyapunov exponent of the GOY shell model
scales as $\hbox{Re}^{\alpha}$ where 
\begin{equation}
\label{eq:alpha1}
 \alpha= \max_{h} \left[\frac{D(h)-1-2h}{1+h}\right]. 
\end{equation}
It is indeed numerically observed to scale 
as $\hbox{Re}^{0.495}$, in good agreement with a computation of $\alpha$ staring from a 
function $D(h)$, obtained by a parametric fit of 
measured values of the scaling exponents $\zeta_q$ in 
experiments\cite{BPPV,Crisanti93}.

The mean square fluctuations at the Kolmogorov scale
are $\langle |u_{n^*}|^2\rangle\sim\hbox{Re}^{-1/2}$.
If we compute the distance between two shell variable configurations as
\begin{equation}
|u-u'| = \sqrt{\sum_n |u_n-u_n'|^{2}},
\label{eq:error_def}
\end{equation}
an error smaller than $O(\hbox{Re}^{-1/4})$ is relatively small all
over the inertial range. It can be taken to be infinitesimal,
and its growth rate will be the fastest linear growth rate.

If, however, the error is larger than $O(\hbox{Re}^{-1/4})$, it could be larger
than the typical size of the fluctuations at the Kolmogorov scale.
Such an error would have to be concentrated on larger scales,
since otherwise we have that for some $n$ $u_n$ and/or $u'_n$ 
is much larger than the typical size.
In other words, a physical perturbation
larger than  $O(\hbox{Re}^{-1/4})$ cannot be obtained by 
a random perturbation of that size uniformly distributed
over all the shells.

In the Appendix \ref{app:A} we discuss some possible definitions for the finite 
Lyapunov-exponents. In what follows we adopt the following procedure: 
after a long integration time to let the system relax towards the statistically 
stationary state, we introduce a very small error. This is done by
generating a new shell variable configuration $u_n'$ differing from
$u_n$ by a small fraction of $\langle |u_n|^2\rangle^{1/2}$.
Another possible approach would be, that if we want an initial error of
size $\epsilon$, we determine a shell $n_{\epsilon}$ such that
$\langle |u_{n_{\epsilon}}|^2\rangle^{1/2} $ is about $\epsilon$, 
and we concentrate the perturbation on shells above $n_{\epsilon}$.

We then iterate $u_n$ and $u_n'$ (perturbed system) for
again a long time, such that the error has grown to a threshold,
which is still small compared to $V\hbox{Re}^{-1/4}$.
We thus have two realizations of configurations in the statistically
stationary state, which only differ by a small error,
which we call $\delta_0$.

Further we define a series of thresholds $\delta_n = r^n\delta_0$,
and we measure the times it takes for the error to grow from
$\delta_0$ to $\delta_1$, and so on.
For brevity we will call
these times error doubling times, even if $r$ can be different from two. 
The threshold rate $r$ should not
be taken too large, because then the error has to grow through
several different scales before reaching the next threshold.
On the other hand, the rate $r$ can not be too close to one, so
a sensible threshold rate is on the order of two.
The most convenient
choice of $r$ clearly depends on the how the fluctuations in the shell
variable depend on
$n$, and in what way we measure the error.
For our model (\ref{eq:sh3}) with error measured
by (\ref{eq:error_def}), the range of error sizes in the inertial
range is not large, scaling as $\hbox{Re}^{-1/4}$. 
For 35 shells, which is the largest systems we simulate, we can take
$\hbox{Re}$ equal to $10^{10}$, which gives an error range of about $300$.
With error threshold rate equal to two, that would give about $8$ data
points, with some points lost on both ends due to boundary effects.
For practical reasons we therefore take $r$ equal to $\sqrt{2}$.

When we have performed $N$ error doubling experiments, we can form
an estimate of the expectation value of some quantity $A$:
\begin{equation}
\langle A \rangle_e = {1\over N} \sum_{i=1}^N A_i.
\end{equation}
This is not the same as taking a time-average, since different error
doubling experiments may take different times. Indeed, we have
\begin{equation}
  \langle A \rangle_t = {1\over T} \int_0^T A(t)\, dt 
   = {{\sum_i A_i\, \tau_i}\over{\sum_i \tau_i}} 
   = {{\langle A\,\tau \rangle_e}\over{\langle \tau\rangle_e}}.
\label{eq:average}
\end{equation}
A particular case of the above relation concerns the mean error doubling
times themselves. Let $T_r(\delta_n)$ be the time it takes for an error to grow
from threshold $\delta_n$ to $\delta_{n+1}$. Then
\begin{equation}
  \lambda(\delta_n) = \left\langle {1\over{T_r(\delta_n)}}\right\rangle_t\,\ln r
                    = {1\over{\langle T_r(\delta_n) \rangle_e}}\,\ln r ,
\label{eq:finite-lyapunov}
\end{equation}
where we have used the definition 
of (\ref{eq:predict}).

The finite-size Lyapunov exponents, $\lambda(\delta_n)$, can be compared
with shell
turn-over times as follows: we first select a shell $n_{\delta}$
such that $\langle |u_{n_{\delta}}|^2 \rangle^{1/2}$ is about
$\delta$, and then
estimate $\tau_{\delta}^{-1}$ as $k_{n_{\delta}} \delta$,
which
scales as $\delta^{-2}$. This argument for typical
error growth times is the same as Lorenz argument for
$3D$ turbulence\cite{Lorenz69}, discussed
above in section \ref{s:multifractal}.

\begin{figure}[hbt]
\epsfbox{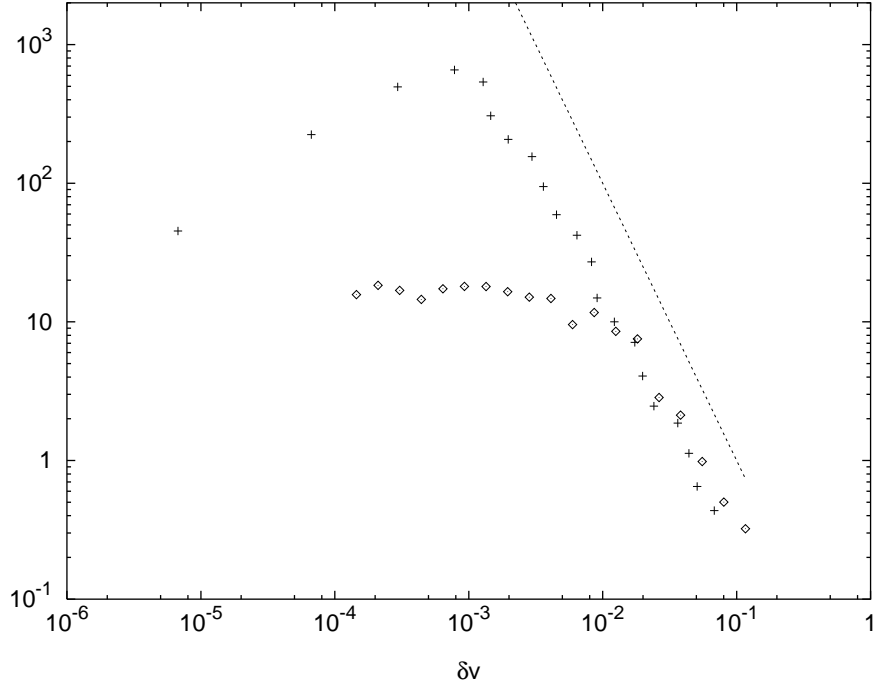}
\vspace{15pt}
\caption{Error doubling times (diamond) compared with shell turn-over times (plus).
         Number $N$ of simulated shells is $27$, and Reynolds number 
         ${\rm Re}= \nu^{-1}= 10^{9}$, $k_0= 0.05$ and $f= (1+i)\times 0.005$.
         The equations were integrated with
         a slaved-frog scheme\protect\cite{Pisarenko93,FrischSheThual},
         with constant time-step $2\cdot 10^{-6}$.
         The initial perturbation
         was randomly uniform over all
         shells in the inertial range,
         with amplitude less than $10^{-6}$.
         The perturbed and unperturbed
         configurations were integrated
         until the error reach the first threshold $\delta_0$ at
         $10^{-4}$.
         The error growth rate parameter $r$ is $2^{1/2}$.
         The number of error doubling experiments was
         $400$. The dashed line has slope $-2$.
        }
\label{f:doubling-times}
\end{figure}

In Fig.\ref{f:doubling-times} we compare error doubling
times and shell turn-over times, as a function of size
of the perturbation and of the typical fluctuations in the
corresponding shell.
Below the Kolmogorov
scale, the turn-over times increase: we are here in the dissipation
range, where the shell amplitudes decrease
quickly. On the other hand, the doubling times tend to a constant
as the error threshold is small. We are here in the infinitesimal
range, and the constant is approximatively
the inverse of the Lyapunov exponent.
At the Kolmogorov scale, there is rather large
discrepancy between the Lyapunov exponent and the 
turn-over time. This observation, that the Lyapunov
exponent obeys a scaling law with a sizeable numerical
pre-factor, has been made before\cite{JensenPaladin91},
but without a plausible explanation.
We here find nice agreement with our prediction that
the inertial range for the finite-size Lyapunov exponent
is shorter than the spectral inertial range, because the
first is limited from below by the scaling exponent $h_3$,
as in equation (\ref{eq:fluctuation-scale}), while the second is limited from
below by the scaling exponent $h_2$\cite{FrischVergassola}.

\begin{figure}[hbt]
\epsfbox{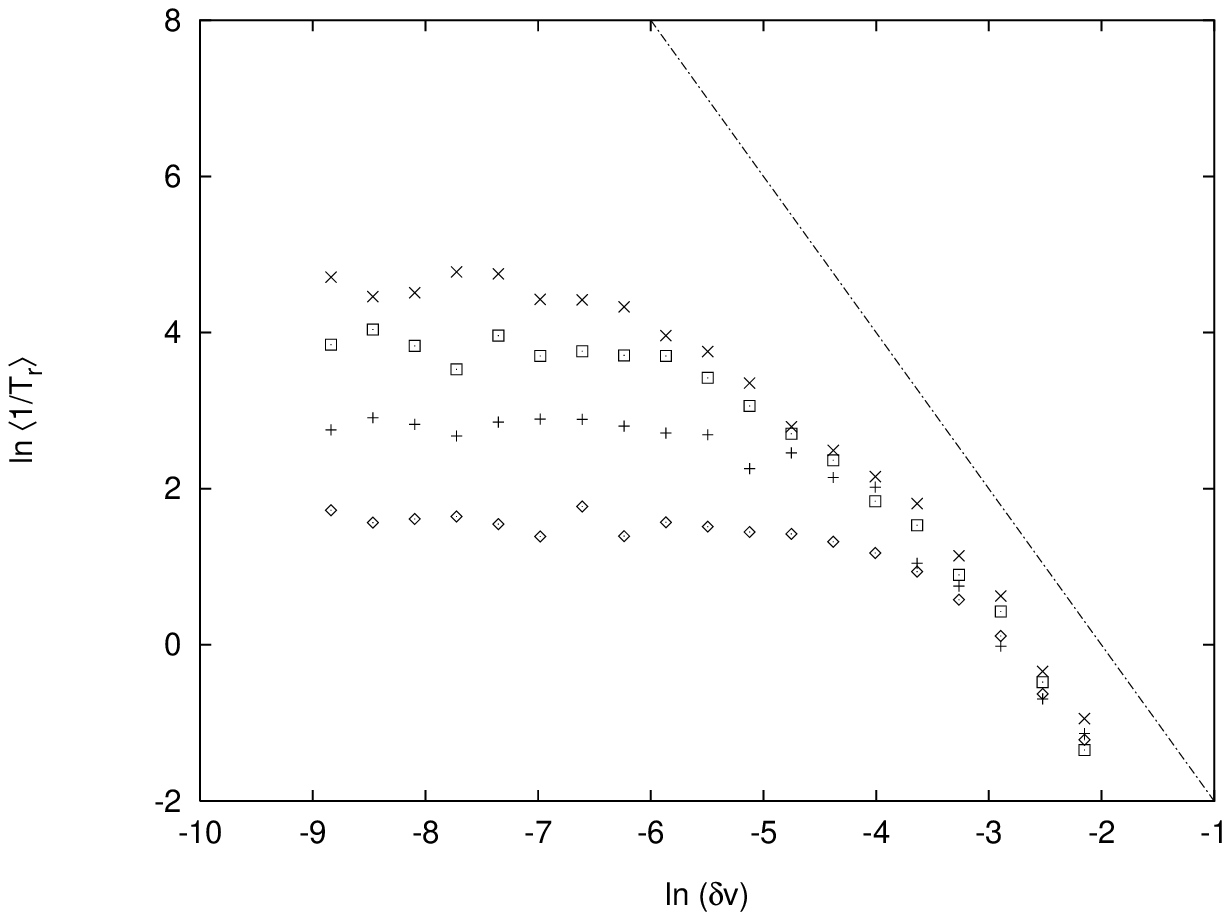}
\vspace{15pt}
\caption{$\ln\langle 1/T_r(\delta v)\rangle$
         versus $\ln\left(\delta v\right)$
         for different Reynolds numbers ${\rm Re}=\nu^{-1}$.
         Parameters as in Fig.\protect\ref{f:doubling-times},
         except that the times-step has been adjusted to the changing
         viscosity. The different symbols refer to:
         $N = 24$ and $\nu=10^{-8}$ (diamond);
         $N = 27$ and $\nu=10^{-9}$ (plus);
         $N = 32$ and $\nu=10^{-10}$ (square);
         $N = 35$ and $\nu=10^{-11}$ (cross).
         The straight line has slope $-2$.
        }
\label{f:Re-scaling}
\end{figure}

\begin{figure}[hbt]
\epsfbox{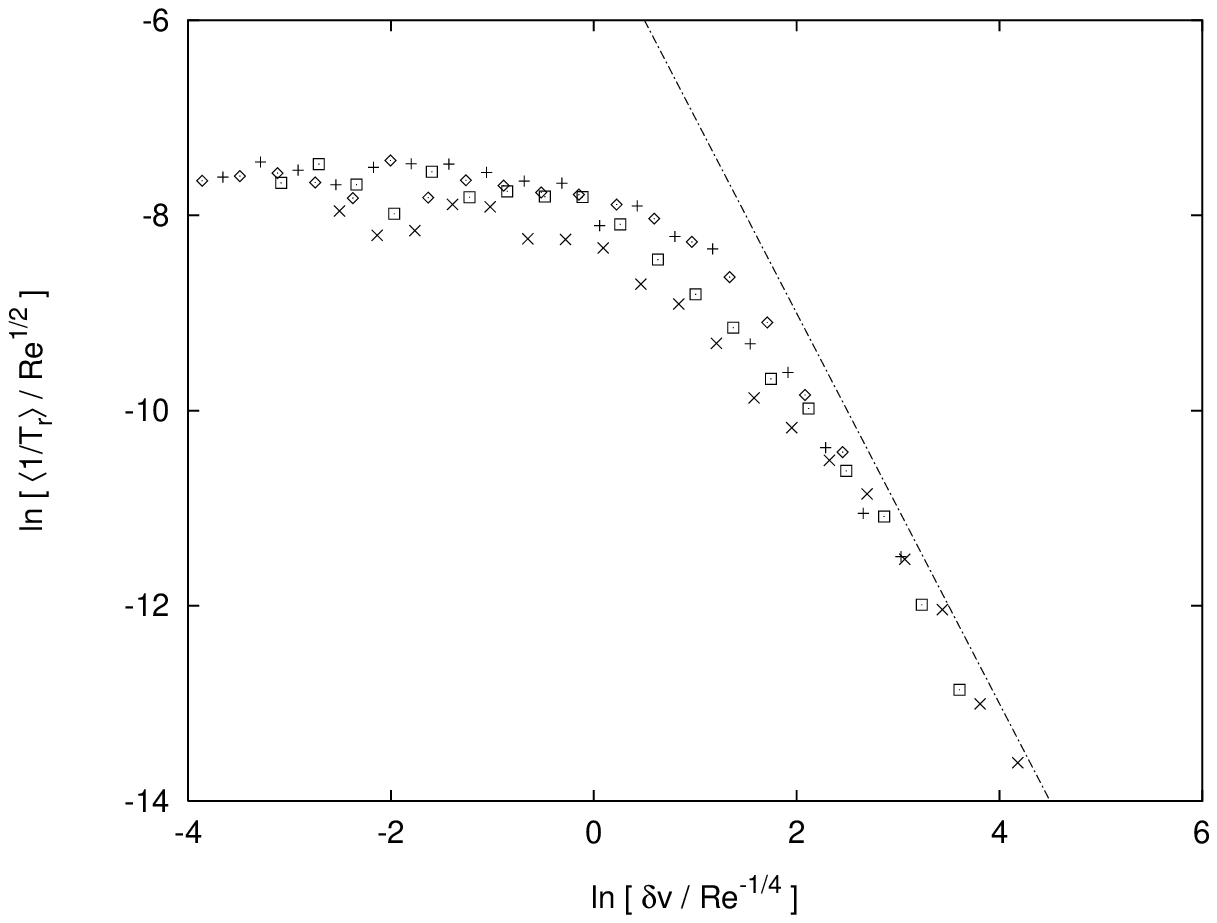}
\vspace{15pt}
\caption{$\ln\left[\langle 1/T_r(\delta v)\rangle/{\rm Re}^{1/2}\right]$
         versus $\ln\left[\delta v/{\rm Re}^{-1/4}\right]$
         at different Reynolds numbers ${\rm Re}=\nu^{-1}$.
         Parameters as in Fig.\protect\ref{f:doubling-times},
         except that the times-step has been adjusted to the changing
         viscosity. The different symbols refer to:
         $N = 24$ and $\nu=10^{-8}$ (diamond);
         $N = 27$ and $\nu=10^{-9}$ (plus);
         $N = 32$ and $\nu=10^{-10}$ (square);
         $N = 35$ and $\nu=10^{-11}$ (cross).
         The straight line has slope $-2$.
        }
\label{f:Re-data-collapse}
\end{figure}

In Fig. \ref{f:Re-scaling} we compare the error doubling
times for different Reynolds numbers. 
For small thresholds the doubling times scale 
as the Lyapunov exponent, i.e. as $\hbox{Re}^{-1/2}$.
We also observe that the bend away from the infinitesimal
growth rate occurs at smaller error scales for larger
Reynolds numbers. This suggests that a simple
scaling ansatz can be sought in the following form:
times and errors are scaled with the turn-over time
and the typical scale of fluctuations at the
Kolmogorov scale, that is by
$\hbox{Re}^{-1/2}$ and
$\hbox{Re}^{-1/4}$, respectively.
In Fig. \ref{f:Re-data-collapse}
we show such re-scaled data. The data collapse
is reasonable. 

\begin{figure}[hbt]
\epsfbox{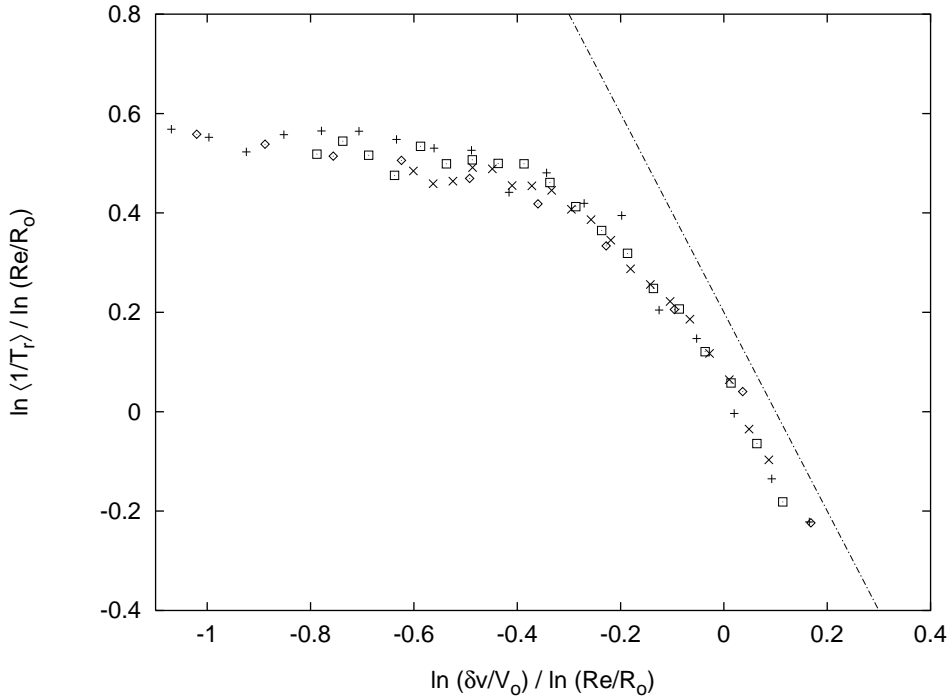}
\vspace{15pt}
\caption{Multiscaling data collapse [see eq. (\protect\ref{eq:multiscaling})].
         Parameters as in Fig.\protect\ref{f:doubling-times},
         except that the times-step has been adjusted to the changing
         viscosity. The different symbols refer to:
	 $N = 24$ and $\nu=10^{-8}$ (diamond);
	 $N = 27$ and $\nu=10^{-9}$ (plus);
	 $N = 32$ and $\nu=10^{-10}$ (square);
	 $N = 35$ and $\nu=10^{-11}$ (cross).
	 The straight line has slope $-2$.
         The fitting parameters are $R_o = 6\times 10^{6}$,
         $V_o=  5\times 10^{-2}$, and $\mbox{Re}= \nu^{-1}$.
        }
\label{fig:multiscaling}
\end{figure}

To improve the data collapse, taking into account multifractal corrections,
we made a scaling based on multiscaling \cite{FrischVergassola}, i.e. 
of the form 
\begin{equation}
\label{eq:multiscaling}
\frac{\ln\langle 1/T_r(\delta v)\rangle }{\ln(\hbox{Re}/\hbox{R}_o) } = 
                             f\bigl(\ln(\delta v/V_o) / \ln(\hbox{Re}/\hbox{R}_o)\bigr)
\end{equation}
where $\hbox{R}_o$, $V_o$ are parameter to be fixed, and $f(x)$ is the scaling function.
According the argument at the end of section \ref{s:multifractal}, we have
$f(x)\sim x^{-2}$ for large $x$, while $f(x)$ is constant for small values of
$x$. In the intermediate regime $f(x)$ has a nontrivial form which depends on the shape of
$D(h)$, as follows from eqs. (\ref{eq:multi-verg})-(\ref{eq:chi1}).
The result is shown in Fig. \ref{fig:multiscaling}. The data collapse is clearly improved.

We conclude this section discussing the case of $2D$ turbulence. 
Two dimensional Euler equation has the peculiar 
property of an infinite number of invariants. 
Two of them are retained in a finite Fourier discretization, 
the energy and the average square vorticity, or enstrophy. 
As we previously discussed, the second
conserved quantity in the GOY shell model depends on the 
choice of the parameter $\epsilon$. With the choice
$\epsilon = 5/4$, leading to
\begin{equation}
a = 1, \qquad b = -\frac{5}{4}, \qquad c = \frac{1}{4},
\label{eq:GOY-2D}
\end{equation}
equations (\ref{eq:sh1}) conserve in the unforced and inviscid limit, 
in addition to the energy, the enstrophy here defined as
\begin{equation}
   Z = {1 \over 2} \sum_n k_n^2\, | u_n |^2.
\label{eq:enstrophy}
\end{equation}

Despite the fact that the two-dimensional shell model has superficially the same
physical justification of its three-dimensional corresponding model, it has
been demonstrated that it has little to do with turbulence 
\cite{Aurell94,DitlevsenMorgensen}. 
Moreover, all the numerical simulations of two-dimensional Navier-Stokes
equation at sufficiently high Reynolds number have demonstrated the
dynamical relevance of coherent structures which emerge spontaneously
from the turbulent flow. The predictability problem, which is more 
relevant for geophysical flows in this case than in $3D$ turbulence, 
is also ruled by coherent vortex motion in the physical space, rather 
than modes dynamics in Fourier space \cite{BCCV96}. 

\begin{figure}[hbt]
\epsfbox{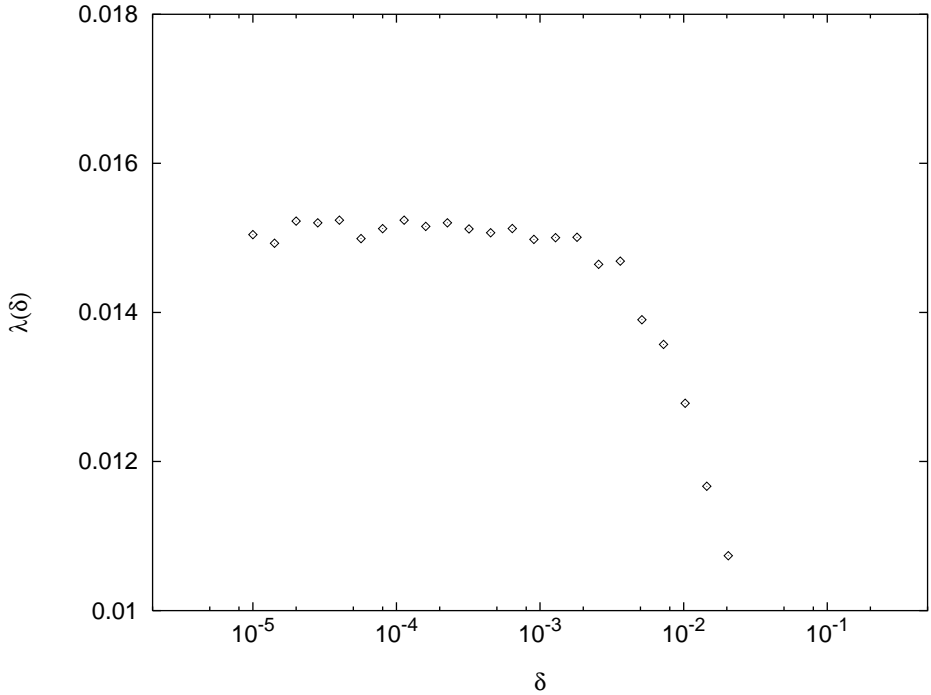}
\vspace{15pt}
\caption{Scale dependent Lyapunov exponent $\lambda(\delta v)$ for 
         two-dimensional shell model with $N=24$ shells, $k_0=0.05$, 
         $\nu = 10^{-15}$ and $f= (1+i)\times 0.005$. 
         Note the linear scales on the ordinate axis.
         }
\label{fig:2d}
\end{figure}

With this limitations, the study of the predictability problem, as
addressed in the present paper, in two-dimensional shell model is
nevertheless interesting because of the different scaling behavior with
respect to the $3D$ situation.
Dimensional analysis \cite{KM80} shows that in the enstrophy cascade one
expects constant -- i.e. independent on the scale -- turn-over times. Hence
an argument similar to that of section \ref{s:multifractal} shows that 
\begin{equation}
  \left\langle {1 \over T_r(\delta v)} \right\rangle \sim \hbox{const}
                                       = \lambda_{\rm max} / \ln r,
\label{eq:pr2d}
\end{equation}
where $\lambda_{\rm max}$ is the largest Lyapunov exponent. 
The predictability time for two-dimensional shell model is thus
determined by a single value $1/\lambda_{\rm max} \sim \epsilon_{Z}^{-1/3}$, 
where $\epsilon_{Z}$ is the enstrophy flux toward small scales, 
up to a perturbation of the order of the large scale velocity field, 
where saturation effects are dominant. 
Figure \ref{fig:2d} shows the finite-size Lyapunov exponent 
$\lambda(\delta v)$ for a simulation with $N=24$ shells. The
forcing term is now $f=5 \times 10^{-4}(1+i)$ and $\nu=10^{-8}$.
Because of the inverse energy cascade we introduce in the equation
(\ref{eq:sh1}) an artificial large-scale dissipation 
$\nu'/k_n$ ($\nu'=10^{-5}$) in the first shells $n \leq 3$
\cite{YamadaOhkitani88a}.

The velocity field shows a pseudo-cascade power law,  see
\cite{Aurell94},  
$\langle |u_n|^2 \rangle^{1/2} \sim \epsilon_{Z}^{1/3} k_n^{-1}$ 
in a wide range $1 \le k_n \le 10^3$. The mean enstrophy 
flux in this range is $\epsilon_{Z}=5 \times 10^{-6}$ which 
is in agreement with the dimensional evaluation of the eddy
turnover time $\tau_{n} \sim \epsilon_{Z}^{1/3}$. 

The data plotted in figure \ref{fig:2d} are obtained by using the 
method described in appendix \ref{app:A} for computing the size dependent
Lyapunov exponent. The same results,  not reported,  can be obtained 
from the doubling time algorithm.

We stress once more that, in the light of the results discussed in
\cite{Aurell94}, two-dimensional shell model is not a good model for
two-dimensional turbulence. As a consequence a discussion of the predictability problem
for two-dimensional turbulence requires the direct study of Navier-Stokes
equations \cite{BCCV96}. Preliminary results show that this scenario
remains, nevertheless, valid in the direct cascade \cite{noi}.

\section{Closure Approximation}
\label{s:EDQNM}
In this Section we describe the results obtained from 
the eddy damped quasi-normal Markovian approximation (EDQNM)
for the shell model. The basic idea of closure approximations 
is quite simple: write down the Reynolds hierarchy for moments
of the shell variables and truncate the chain to the lowest
sensible order. 
The important point is that in the closure approximation
intermittent effects are washed out, so we can directly test 
if the relevant mechanism is due to the existence of many
characteristic times. We do not report the derivation of
the EDQNM equations for the shell model. The interested reader can 
find it in Ref. \cite{Aurell96}.

We consider two independent realizations of the shell model 
field, $u_n$ and $v_n$, with the same energy spectrum
$E_n = \langle u_n\,u_n^*\rangle = \langle v_n\,v_n^*\rangle$, and both 
evolving according the shell model equations 
(\ref{eq:sh1})-(\ref{eq:GOY_standard}).
The distance between the two fields can be obtained from, cfr. (\ref{eq:error_def}),
the energy difference at shell $n$:
\begin{equation}
\label{eq:Dn}
  \Delta_n = \frac{1}{2}\,\langle (u_n - v_n)\,(u_n^* - v_n^*) \rangle
           = (E_n - \Re\, W_n),
\end{equation}
where $W_n=\langle u_n\,v_n^* \rangle$, and $\Re$ denotes the real part.
. From the definition it follows
\begin{equation}
\label{eq:EDQNMdist}
 \delta v(t) = \left[\sum_n \Delta_n(t)\right]^{1/2}.
\end{equation}

The evolution equations of $E_n$ and $W_n$ in the EDQNM approximation
read:
\begin{equation}
\label{eq:EnEDQNM}
  \begin{array}{rl}
         \left(\frac{d}{dt} + 2\nu k_n^2\right)\, E_n = 2&\Bigl[
                k_n^2\,    \theta(n,t)\, 
 (E_{n+1}\,E_{n+2} - {1 \over 2}\,E_n\,E_{n+2} - {1 \over 2}\,E_n\,E_{n+1}) \\ 
       &\,-{1 \over 2}\,k_{n-1}^2\,\theta(n-1,t)\, 
 (E_{n}\,E_{n+1} - {1 \over 2}\,E_{n-1}\,E_{n+1} - {1 \over 2}\,E_{n-1}\,E_n) \\
       &\,-{1 \over 2}\,k_{n-2}^2\,\theta(n-2,t)\, 
 (E_{n-1}\,E_n - {1 \over 2}\,E_{n-2}\,E_n - {1 \over 2}\,E_{n-2}\,E_{n-1})
                                                          \Bigr] \\ 
       &\,+ 2\,\epsilon\,\delta_{n,4},
  \end{array}
\end{equation}
and
\begin{equation}
\label{eq:WnEDQNM}
 \begin{array}{rl}
 \left(\frac{d}{dt} + 2\nu k_n^2\right)\, W_n = 2&\Bigl[
 k_n^2\,    \theta(n,t)\, 
 (W_{n+1}^*\,W_{n+2}^* - {1 \over 2}\,W_n\,E_{n+2} - {1 \over 2}\,W_n\,E_{n+1})
 \\ &\,-
 {1 \over 2}\,k_{n-1}^2\,\theta(n-1,t)\, 
 (W_{n}\,E_{n+1} - {1 \over 2}\,W_{n-1}^*\,W_{n+1}^* - {1 \over 2}\,E_{n-1}\,W_n)
 \\ &\,-
 {1 \over 2}\,k_{n-2}^2\,\theta(n-2,t)\, 
 (E_{n-1}\,W_n - {1 \over 2}\,E_{n-2}\,W_n - {1 \over 2}\,W_{n-2}^*\,W_{n-1}^*)
 \Bigr] \\ &\,+
 2\,\epsilon\,\delta_{n,4},
 \end{array}
\end{equation}
where
\begin{equation}
  \theta(n,t)= \frac{\displaystyle 1 - e^{-[ \nu (k_n^2 + k_{n+1}^2 
                                                + k_{n+2}^2 ) +
                      \mu_n + \mu_{n+1} + \mu_{n+2} ]\,t}
                }
                {\displaystyle 
                     \nu (k_n^2 + k_{n+1}^2 + k_{n+2}^2 ) +
                      \mu_n + \mu_{n+1} + \mu_{n+2}
                  }, 
        \label{eq:theta}
\end{equation}                                         
and
\begin{equation}
\label{eq:mu}
 \mu_n \equiv \mu(k_n,E_n) = \alpha\,k_n\,E_n^{1/2}.
\end{equation}
We have one free parameter, the dimensionless constant
$\alpha$. It should be adjusted such that the spectrum is as similar as
possible to the spectrum obtained in simulations of the full equation.
The energy spectrum of the shell model in the
EDQNM approximation must therefore obey $E_n \simeq C(\alpha)
\epsilon^{2/3} k_n^{-2/3}$ in the inertial range. The undetermined
function $C(\alpha)$ is the Kolmogorov constant.

On the other hand it has become clear in several independent
investigations that intermittency corrections exist in shell models. The
energy spectrum is therefore in reality more closely described by $E_n
\sim F(\epsilon)k_n^{-\zeta_2}$, where the exponent $\zeta_2$ has been
estimated to be $0.70$ \cite{JensenPaladin91}. The function
$F$ that gives the prefactor to the power law in the inertial range
should not depend on viscosity, but depends on the forcing
through $\epsilon$, the mean dissipation of energy per unit time, or,
equivalently, the mean energy input into the system from the force. In a
really large inertial range the two power-laws are not good
approximations to one other. The best that can be done is to demand that
the spectra agree as closely as possible at the upper end of the
inertial range.  A reasonable agreement is obtained for
$\alpha=0.06$, leading to $C(\alpha)= 1.5$ which is the value 
observed both in simulations of the shell model and in experiments
\cite{Orszag}.

The procedure described in the previous Section to compute the scale dependent
Lyapunov exponent for the shell model can be adopted here for the 
closure equations. In practice after a long iteration time, to have a 
well stabilized energy spectrum $E_n$, we take a small initial distance
$\delta v(0)$ and perform the doubling experiment similar to those of
the previous section iterating equations (\ref{eq:WnEDQNM}). 

\begin{figure}[hbt]
\epsfbox{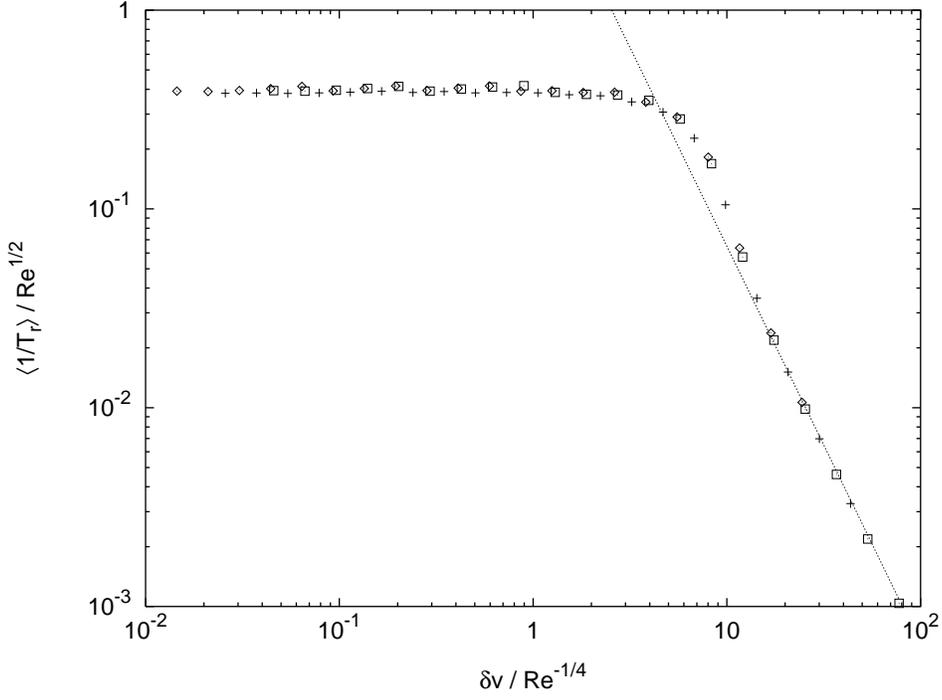}
\vspace{15pt}
\caption{$\langle 1/T_r(\delta v)\rangle / {\rm Re}^{1/2}$ as a function of
         $\delta v / {\rm Re}^{-1/4}$ for different values of the Reynolds
         number ${\rm Re} = \nu^{-1}$ for the EDQNM approximation. The different
         symbols refer to: $\nu = 10^{-8}$ (diamond), $\nu = 10^{-9}$ (plus) and
         $\nu = 10^{-10}$ (square). The line gives the Lorenz results $-2$. The
         other parameters are $N=32$, $k_0=0.05$, $\epsilon = 1$ and $\alpha = 0.06$.
        }
\label{fig:edqnmfig1}
\end{figure}

In Fig. \ref{fig:edqnmfig1} we show $\langle 1/T_r(\delta v)\rangle / {\rm Re}^{1/2}$
as a function of the rescaled distance $\delta v/{\rm Re}^{-14}$, for different
${\rm Re} = \nu^{-1}$. The other parameters are $N=32$ shells, $k_0=0.05$,
integration step $10^{-6}$ and $r=2^{1/2}$. From Fig. \ref{fig:edqnmfig1} we see
that the closure approximation leads to the same scenario observed for
the shell model, confirming that this is due to the existence of many characteristic 
scales. We note that the slope of the curve for $\delta v / {\rm Re}^{-1/4} > 10$ is
the Lorenz value $-2$ since in the EDQNM approximation there are not 
intermittent corrections.

\begin{figure}[hbt]
\epsfbox{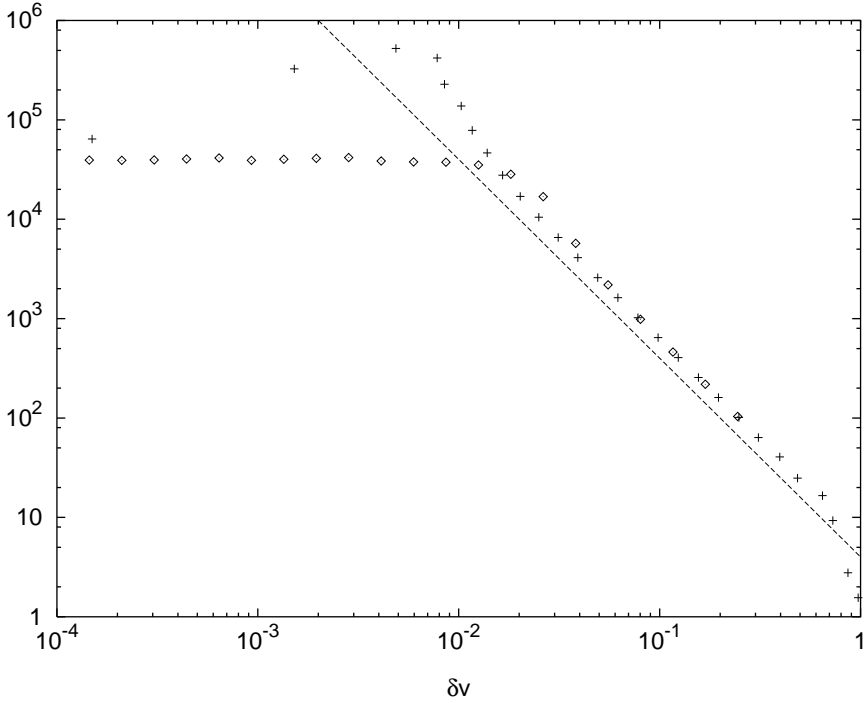}
\vspace{15pt}
\caption{$\langle 1/T_r(\delta v)\rangle$ (diamond) as a function of $\delta v$
         for the EDQNM approximation of GOY model
         with  $N=32$, $k_0= 0.05$, $\epsilon=1$ and $\nu=10^{-10}$.
         The plus are the inverse of the eddy turn-over times
         $\tau^{-1}(\delta v)=k_n \,  E_n^{1/2}$ 
         versus $\delta v= (2NE_n)^{1/2}$.
         The straight line has slope $-2$.
        }
\label{fig:edqnmfig2}
\end{figure}

We note that in this case, since there is not intermittency, the effective inertial
range roughly coincides with the inertial range. In Fig. \ref{fig:edqnmfig2}
we compare $\langle 1/T_r(\delta v)\rangle$ and the inverse of the
turnover time $\tau^{-1}(n) = k_n E_n^{1/2}$ as a function of the distance
$\delta v$. In the figure we used $r=2$. 

\section{ Conclusion}
\label{s:conclusion}
We have introduced a generalization $\lambda(\delta)$ of the 
leading Lyapunov exponent to finite perturbations of size $\delta$.
Unlike the Lyapunov exponent, $\lambda(\delta)$ contains direct
information on the predictability time for an extended chaotic system
and it is particularly useful in presence of several characteristic
times. Moreover, it is computationally no more expensive than the
standard Lyapunov exponent. 

In the limit of infinitesimal perturbation $\delta$, the finite-size 
Lyapunov exponent gives the leading Lyapunov exponent. The way in which
this limit is reached depends on the details of the particular system
and gives informations about the characteristic time scales. In the case
of $3D$ fully developed turbulence we have found an universal scaling
law $\lambda(\delta) \sim \delta^{-2}$ where the value  $-2$ 
 of the exponent is an invariant of the multifractal approach. 

The scaling law is confirmed by extensive numerical simulations on 
shell model for turbulence at very high Reynolds numbers, 
but we warn that it would be very difficult to observe such a scaling 
in experimental data because of the reduction of the scaling range in
presence of intermittency. 

We have also shown and discussed the relationship between our approach
and the $\epsilon$-entropy results in literature.
In conclusion, we think that it would be useful to compute the
finite-scale Lyapunov exponent in other complex dynamical systems, as a
new tool for investigating the presence of characteristic times and the
predictability properties.

\section{Acknowledgments}
This work was supported by the Swedish Natural Science
Research Council through grant S-FO-1778-302 (E.A.),
by INFN ({\it Iniziativa Specifica Meccanica 
Statistica FI11}) and by CNR (Progetto Coordinato {\it Predicibilit\'a in
turbolenza geofisica}).
E.A. thanks Dipartimento di Fisica, Universit\`a di Roma
``La Sapienza'' for hospitality.
G. B. thanks the ``Istituto di Cosmogeofisica del CNR'', Torino, for
hospitality.

\appendix 
\section{Finite size Lyapunov exponents and $\epsilon$-entropy}
\label{app:A}
Here we discuss some alternative ways of computing the finite-size
Lyapunov exponent beyond the definition (\ref{eq:predict}). 

The first method is a modification of
the standard technique \cite{BGGS80,WSSV85}. 
We integrate two trajectories $\bbox{u}(t)$ and $\bbox{u}'(t)$
with initial Euclidean distance $\delta u(0)=\delta$ until their
separation becomes larger, at a given time $T_r$ than 
$r \delta$ where $r$ is a given constant coefficient. 
The perturbed trajectory $\bbox{u}'(T_r)$ is then rescaled 
at the original distance $\delta$, keeping the direction 
$\bbox{u}'-\bbox{u}$ constant.
The possible problem with this definition is that it assumes that the
statistically stationary state of the system is homogeneous with respect
to perturbations of finite size. One may plausibly argue that the structure
of the attractor in phase space on which the motion takes place may be
fractal, and not at all equally dense at all distances from a given point.
The procedure outlined would then not necessarily sample in a faithful way
the motion on the attractor.
In practice, as it is showed by the numerical experiments, we do not
find any difference with  the numerical method presented
in the main text, so the effect must be quite small.
For the usual Lyapunov exponent the problem does not exist since we only
use small finite perturbations as an approximation to infinitesimal perturbations.

The finite-size
Lyapunov exponent at scale $\delta$ is  obtained by averaging 
the divergence rate
\begin{equation}
  \lambda(\delta) = \left\langle {1 \over T_r} \ln\left( {\delta u(T_r) 
                              \over \delta u(0)}\right) \right\rangle
                  = \frac{1}{\langle T_r \rangle_e}\,\ln r,
\label{eq:alterna1}
\end{equation}
along the unperturbed trajectory $\bbox{u}(t)$. 
The average $\langle\cdots\rangle_e$ is over many error-doubling experiments.

In the case of infinitesimal error $\delta$, and not too large
factor $r$,  this definition leads to the maximal Lyapunov
exponent $\lambda_{\rm max}$. 

For maps,  the above methods have to be slightly modified since the
distance between two states is not continuous in time. 
 From eq. (\ref{eq:finite-lyapunov}) readily follows
\begin{equation}
\label{eq:app2}
  \lambda(\delta) = \frac{1}{\langle T_r\rangle_e}\, 
                     \left\langle \ln\left( \frac{\delta(T_r)}{\delta}\right)
                     \right\rangle_e,
\end{equation}
where $T_r$ is the minimum time such that the distance between two realizations
is larger or equal to  $r\delta$ and $\delta(T_r)$ is the distance at that time.

Definition (\ref{eq:alterna1}) is valid for any value of $r$, 
which however is 
generally assumed not too large to avoid the interference of different scales. 
In particular, one could think of removing the threshold condition used for 
defining $T_r$ and simply compute the average error growth rate at every time
step. In other words, at every time step $\delta t$ in the integration, the
perturbed trajectory $\bbox{u}'(t)$ is rescaled to the original distance 
$\delta$, keeping the direction $\bbox{u}-\bbox{u}'$ constant.
The finite-size Lyapunov exponent is given by the average of the one-step 
exponential divergence:
\begin{equation}
\label{eq:app3n}
  \lambda(\delta) = \frac{1}{\delta t}\, \left\langle
                    \ln\left(\frac{\delta u(t+\delta t)}{\delta u(t)}
                       \right)
                    \right\rangle_{t},
\end{equation}
which is equivalent to the above definition (\ref{eq:alterna1}). This 
method is indeed the one used for computing $\lambda(\delta)$ in the  case of
$2D$ shell model.
The one-step method (\ref{eq:app3n}) has the advantage that it can be easily 
generalized to compute the sub-leading finite-size Lyapunov exponent following the 
standard orthonormalization method \cite{BGGS80}.
One introduces $k$ perturbed trajectories
$\bbox{u}^{(1)},\ldots,\bbox{u}^{(k)}$
each at distance $\delta$ from $\bbox{u}$ and such that
$\bbox{u}^{(k)}-\bbox{u}$ are orthogonal each to the others.
At every time step, any difference $\bbox{u}^{(k)}-\bbox{u}$ is
rescaled at the original value and orthogonalized, while the corresponding
finite size Lyapunov exponent is accumulated according to (\ref{eq:app3n}).
Here we have again the problem of the implicitly assumed homogeneity of
the attractor, but also a problem of isotropy when we re-orthogonalize
the perturbations. We note that this could be a more serious problem, which
will not be discussed here any further.

For systems with only one positive Lyapunov exponent the size dependent 
Lyapunov 
exponent $\lambda(\delta)$, definition (\ref{eq:app2}), 
or (\ref{eq:predict}), coincides with the $\epsilon$-entropy widely described 
below in Appendix \ref{s:kolmogorov} and by
Gaspard and Wang \cite{GaspardWang92,GaspardWang93}. We consider now few examples.

\subsection{Map with noise}
Consider a chaotic deterministic map perturbed by a Gaussian term:
\begin{equation}
\label{eq:app3}
 x(n+1) = f(x(n)) + \sigma y(n),
\end{equation}
where $y(n)$ is a stochastic variable with Gaussian distribution of 
zero mean and unit variance. The $y(n)$ for different times $n$ are independent.
When $\sigma=0$ the map $f(x)$ is chaotic with positive Lyapunov exponent 
$\lambda$. A simple computation shows that
\begin{equation}
\label{eq:app4}
 \delta x(n) \sim \delta x(0)\, e^{\lambda\,n} + \sigma\, \delta y(n-1).
\end{equation}
For $|\delta x(0)| \ll \sigma$ the noise term is negligible, so that 
$\lambda(|\delta x(0)|)\simeq \lambda$. In the opposite limit, 
$|\delta x(0)| \gg \sigma$, the first term in (\ref{eq:app4}) can be neglected so
that $\delta x(n)\sim \sigma\, \delta y(n-1)$. Thus in one iteration of the map
the distance between the two trajectories grows to $O(\sigma)$, larger than the
tolerance. From (\ref{eq:app2}) we have 
$\lambda(|\delta x(0)|)\sim \ln (\sigma/|\delta x(0)|)$. These are the same result
obtained for the $\epsilon$-entropy, see Sect. 3.5 of Ref. \cite{GaspardWang93}.

\subsection{Ornstein-Uhlenbeck process and Yaglom noises}
Consider the Gaussian process described by the Langevin equation
\begin{equation}
\label{eq:app5}
 \frac{dx}{dt} = -ax + c\eta,
\end{equation}
where $a>0$ and $\eta$ is a white noise with zero mean and correlation
$\langle\eta(t)\,\eta(t')\rangle_{\eta} = \delta(t-t')$. The formal solution
of (\ref{eq:app5}) reads
\begin{equation}
\label{eq:app6}
 x(t) = e^{-at} x(0) + c\int_{0}^{t} e^{-a(t-t')}\, \eta(t')\, dt',
\end{equation}
so that the distance between two different process behaves in time as
\begin{equation}
\label{eq:app7}
 \delta x(t) = e^{-at} \delta x(0) + c\int_{0}^{t} e^{-a(t-t')}\, 
                     \delta\eta(t')\, dt'.
\end{equation}
This implies that $\delta x(t)\sim \sqrt{ct}$, and thus the predictability time
$T(\delta x(0), \Delta)$ behaves as $T(\delta x(0), \Delta) \sim (\Delta/c)^2$ so that
$\lambda(\delta x(0), \Delta) \sim (c/\Delta)^2 \sim (c/\delta x(0))^2$, as found for
the $\epsilon$-entropy, see
below or Sect.~3.6.2 of \cite{GaspardWang93}.

Similar computations can be done for the Yaglom noise, see Sect.~3.6.3 of 
\cite{GaspardWang93}.

\subsection{Model of deterministic diffusion}
\begin{figure}[hbt]
\epsfbox{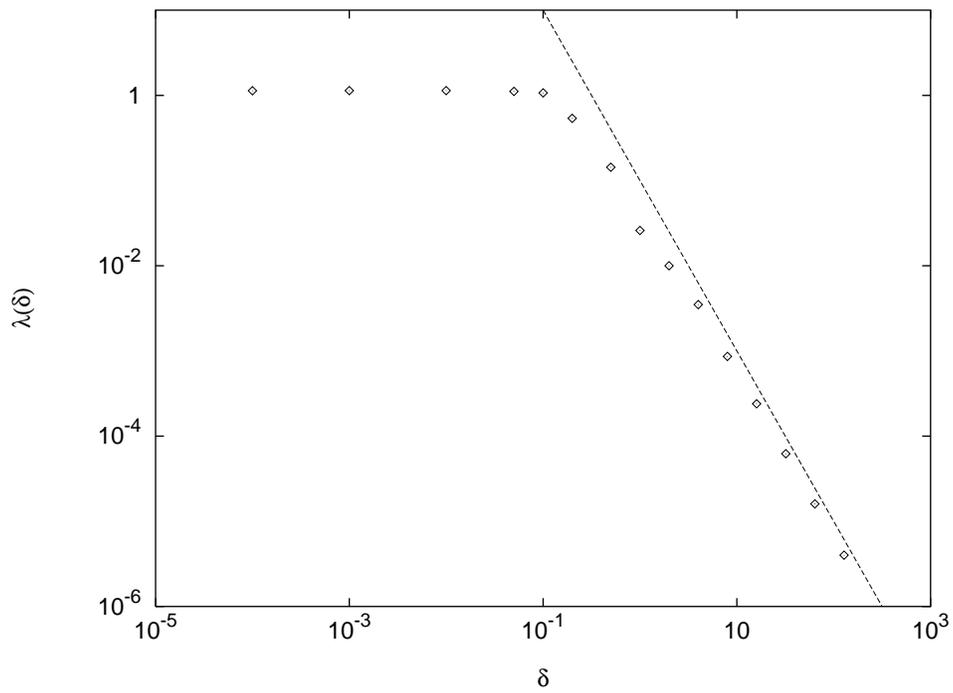}
\vspace{15pt}
\caption{Scale dependent Lyapunov exponent $\lambda(\delta)$ for 
         the map (\protect\ref{eq:app8}). The line has slope $-2$.
        }
\label{fig:det-diff}
\end{figure}

The one-dimensional map
\begin{equation}
\label{eq:app8}
 x(n+1) = x(n) + p\sin(2\pi x(n)),
\end{equation}
presents deterministic diffusion. For example for $p=0.8$ the diffusion coefficient
is $D\simeq 0.18$. 
The size dependent Lyapunov exponent for this map can be computed numerically using the
definition (\ref{eq:app2}). From Fig. \ref{fig:det-diff} we see that
$\lambda(\delta) \simeq \lambda $ for small $\delta$, 
while $\lambda(\delta)\sim \delta^{-2}$ for large values of $\delta$. The 
$\epsilon$-entropy shows the same behavior, see Fig. 25.b of Ref. \cite{GaspardWang93}. 
Note that in Ref. \cite{GaspardWang93} the $\epsilon$-entropy is measured in 
unit of digit/iteration, so there is a multiplicative factor $\ln 10 \simeq 2.3$
with $\lambda(\delta)$ of Fig. \ref{fig:det-diff}.

\section{The Kolmogorov result on $\epsilon$-entropies of
Gaussian processes}
\label{s:kolmogorov}
The purpose of this appendix is to derive systematically
on a physical level of rigor the results of
Kolmogorov on the $\epsilon$-entropy of Gaussian variables
and Gaussian processes and random fields.
These results were stated without proof by Kolmogorov \cite{Kolmogorov56},
and in the review by Tikhomirov \cite{Tikhomirov}.
The published proofs we are aware of are either not easily accessible,
or carried out in such generality that the simple underlying idea may
be lost to the less mathematically inclined reader. Hence the interest
of including  a simple derivation here. Let us just
for completeness refer to comparatively recent paper \cite{PinskerSofman}.

The general setting is that of
two random variables $\xi$ and $\eta$, taking values in
spaces $X$ and $Y$. The values in two realizations of
$\xi$ and $\eta$ are denoted $x$ and $y$, respectively, where $x$ lies
in $X$ and $y$ lies in $Y$. 
When $X$ and $Y$ are continuous the
probability distributions of $\xi$ and
$\eta$ will be denoted $P_{\xi}(dx)$ and $P_{\eta}(dy)$.
Except for the trivial case, the
 variables $\xi$ and $\eta$ are not independent, but characterized by
the joint probability distribution $P_{\xi, \eta}(dx,dy)$.

The single-variable probabilities are determined from the joint
distribution as
\begin{eqnarray}
P_{\xi}(dx) &=& \int_y P_{\xi, \eta}(dx,dy) \qquad
P_{\eta}(dy) = \int_x P_{\xi, \eta}(dx,dy)
\label{eq:single}
\end{eqnarray}

The conditional probabilities are 
\begin{eqnarray}
P_{\xi | \eta}(dx) &=& {{P_{\xi, \eta}(dx,dy)}\over{P_{\eta}(dy)}} \qquad
P_{\eta | \xi}(dy) = {{P_{\xi, \eta}(dx,dy)}\over{P_{\xi}(dx)}} 
\label{eq:conditional}
\end{eqnarray}

\subsection{Preliminaries from information theory}
\label{ss:preliminaries}

We suppose in this subsection that
the spaces $X$ and $Y$ are discrete, and consist of finitely
many points, $x_1,\ldots,x_n$ and $y_1,\ldots,y_m$, and associated probabilities
$p_1,\ldots, p_n$  and $q_1,\ldots, q_m$.

The {\it entropy} of the random variable $\xi$ is 
\begin{equation}
H(\xi) = - \sum_{i=1}^n p_i \ln p_i 
\label{eq:entropy}
\end{equation}
If we want to code a message of
$N$ letters, which are given by $N$ consecutive independent
realizations of $\xi$, so
taking values in $x_1,\ldots x_n$, 
this can be done,
in the limit when $N$ is large, by using  $H(\xi)/\ln 2$
bits per letter \cite{Shannon49}.
The pair $\xi$ and $\eta$ specify both a source signal and the output
after sending the signal through
a channel of transmission.
The joint probability distribution $P_{\xi, \eta}$  can  be decomposed
either to the pair $\xi$ (an input)
 and $\eta | \xi$ (an output, given the input),
or to the pair $\eta$ and $\xi | \eta$.
In the discrete case these
second conditional probabilities are denoted
$p_{i | j}$, where $i$ ranges from one to $n$, and $j$ ranges from
one to $m$.
The {\it equivocation} is 
\begin{equation}
\langle H(\xi | \eta)\rangle_y = 
                   \sum_{j} p_j\left( - \sum_{i} p_{i|j} \ln p_{i|j} \right).
\label{eq:equivocation}
\end{equation}
Shannon \cite{Shannon49} considered a {\it gedanken} experiment where we 
send an error-correcting message
parallel to the transmission of $\xi$ to $\eta$, and asked for the number
of bits in the error-correcting message needed to transmit $N$ letters in the
original message virtually without error.
In the limit of large $N$, the answer is $\langle H(\xi | \eta)\rangle_y/\ln 2$
bits per letter.

Suppose further that we have some source of information which we can
recode into letters from $\xi$, then transmit to $\eta$, and then observe.
The rate of transmission of information is not the source
entropy, since we must correct for the equivocation, but the
{\it mutual information}:
\begin{equation}
I(\xi,\eta) = H(\xi) - \langle H(\xi | \eta)\rangle_y. 
\label{eq:mutual-information}
\end{equation}
If we introduce the discrete joint probabilities $p_{i,j}$ and rearrange terms,
we see that (\ref{eq:mutual-information}) can be rewritten as 
in a more symmetrical way, namely:
\begin{equation}
I(\xi,\eta) = \sum_{i,j} p_{i,j} \ln {{p_{i,j}}\over{p_i p_j}}
\label{eq:mutual-information-symmetrized}
\end{equation}

A channel of transmission of information can be considered as a 
collection of pairs $\xi$ and $\eta$, where the $\xi$'s are
the possible inputs, and the $\eta$'s the corresponding outputs.
The {\it capacity} of the channel is the maximum rate of information
transfer:
\begin{equation}
C = \max_{\xi,\eta}\,  I(\xi,\eta) 
\label{eq:capacity}
\end{equation}
where the maximization is performed over the collection of pairs
$\xi$ and $\eta$ that describe the channel.
The fundamental Shannon theorem says that a source signal can be transmitted
over a channel, if the source entropy is less than the channel capacity.

Suppose finally that we have an input
$\xi$ and we wish that the output $\eta$ contains only some partial
information about $\xi$, saying that some
fidelity criterion is fulfilled.
We consider all channels  consisting of the fixed
source $\xi$ and different outputs $\eta$.
The least rate of information transfer needed to specify
$\xi$ in this way is
{\it source entropy with respect to the fidelity criterion}:
\begin{equation}
R = \min_{\eta}\,  I(\xi,\eta) \qquad\hbox{$P_{\xi}$ fixed}
\label{eq:source-entropy-Shannon}
\end{equation}
where the minimization is performed over all $\eta$
which satisfy the criterion with $\xi$.

In practice many fidelity criteria can be written
\begin{equation}
G(\xi,\eta) \leq \epsilon,
\end{equation}
where $G$ is some suitable function and $\epsilon$ is a measure
of the fidelity.
In this case
the source entropy is naturally considered as a function of $\epsilon$,
and we have the Kolmogorov $\epsilon$-entropy
\begin{equation}
H(\xi,\epsilon) = \min_{\eta}\,  I(\xi,\eta), \quad\hbox{where}\quad G(\xi,\eta) \leq \epsilon
\quad\hbox{and $P_{\xi}$ fixed}.
\label{eq:source-entropy-Kolmogorov}
\end{equation}

\subsection{Continuous random variables}
\label{ss:continuous}
Suppose we have a random variable with a continuous distribution, that is
$P_{\xi}(dx) = p_{\xi}(x)dx$ with continuous density $p_{\xi}(x)$.
Then (\ref{eq:entropy}) is
infinite.

As stressed by Kolmogorov, the mutual information, (\ref{eq:mutual-information}),
the capacity, (\ref{eq:capacity}), and, in particular,
 the rate of information production with
respect to a fidelity criterion, 
(\ref{eq:source-entropy-Shannon},\ref{eq:source-entropy-Kolmogorov}),
are all well-defined also in the continuous case.
In other words, although a real number observed with infinite accuracy
contains an infinite amount of information, a real number observed
with finite accuracy contains only a finite amount of
information.

We can see this in a simple heuristic way as follows:
if we introduce a discretization of the
space $X$ into boxes with diameter
$\delta$, we have a new random variable that we
can call $\xi_{\delta}$ with entropy
\begin{equation}
H(\xi_{\delta}) = d\ln\left({1\over{\delta}}\right) - 
         \int p_{\xi}(x)\ln p_{\xi}(x) dx + {\cal O}(\delta),
\label{eq:entropy-discretized}
\end{equation}
where $d$ is the dimension of $X$.

Similarly, from the conditional variable $\xi | \eta$ we have a new
random variable $\xi_{\delta} | \eta$ with equivocation
\begin{equation}
\langle H(\xi_{\delta} | \eta)\rangle_y = d\ln\left({1\over{\delta}}\right) - 
\int p_{\eta}(y)\left( \int p_{\xi | \eta}(x,y)\ln p_{\xi | \eta}(x,y) dx \right) dy +
{\cal O}(\delta).
\label{eq:equivocation-discretized}
\end{equation}
Rearranging terms as in (\ref{eq:mutual-information-symmetrized}),
the mutual information between $\xi_{\delta}$ and $\eta$ is thus
\begin{equation}
I(\xi_{\delta}, \eta) = 
\int  p_{\xi,\eta}(x,y)\, 
   \ln\left({{ p_{\xi, \eta}(x,y)}\over{p_{\xi}(x) p_{\eta}(y)}}\right)\, dx dy +
{\cal O}(\delta).
\end{equation}
As the discretization tends to zero the mutual information tends to a finite value,
provided that the joint probability $P_{\xi, \eta}(dx,dy)$ is not singular with respect
to the product $P_{\xi}(dx) P_{\eta}(dy)$, a result due to Yaglom and Gel'fand\cite{Kolmogorov56}.

We can give a meaning to the most random continuous distribution with respect some
given constraints by
maximizing the entropy of a discretized continuous variable,
as in (\ref{eq:entropy-discretized}), subject to these constraints,
and then letting the discretization tend to zero.
The most random distribution with given first and second moments is thus,
not surprisingly, a Gaussian distribution with the same first and second
moments.
The entropy of a  discretization of a
$d$-dimensional real Gaussian random variable $\xi$ with correlation matrix $C$ is
\begin{equation}
H(\xi_{\delta}) = 
d\ln \left({1\over{\delta}}\right) +
\ln\sqrt{(2\pi e)^d \det C}
+{\cal O}(\delta).
\end{equation}
All  these results on the Gaussian distributions are due to Shannon\cite{Shannon49}.

\subsection{The $\epsilon$-entropy result}
\label{ss:epsilon-entropy}
The spaces $X$ and $Y$ are now identical.
We consider first a one-dimensional Gaussian random variable,
and then a finite-dimensional variable and
show that the  Kolmogorov formula holds in these cases.
In the next subsection, 
we then observe that the Fourier components of Gaussian processes and Gaussian random
fields are independent Gaussian random variables, and so, taking the
the formal infinite-dimensional limit, we find
the desired result.

The fidelity criterion will be the mean square deviation:
\begin{equation}
 \langle(\xi-\eta)^2\rangle \leq \epsilon^2
\end{equation}
Let us first estimate the answer by dimensional arguments.
If the finite-dimensional normal  variable $\xi$ is decomposed
in its principal components, a fluctuation in direction $i$
will be of typical size $\sigma_i$. To estimate the fluctuation
in direction $i$ with an accuracy $\epsilon$ we need about
$\ln (\sigma_i/\epsilon)$ bits.
The answer should therefore be of the form
\begin{equation}
H(\xi,\epsilon) = A \sum_{\sigma_i > \epsilon}
                \ln\left({{\sigma_i}\over{\epsilon}}\right) + B.
\end{equation}
The result of the analysis will be that $A$ is one and $B$ is zero.

Let $\xi$ be a Gaussian one-dimensional random variable
with variance $\sigma^2$.
The random variable $\xi | \eta$ is  defined on the support of $P_{\eta}$.
We suppose it to have mean 
$m(y)$ and variance $\alpha(y)$. 
We have the obvious inequality
\begin{equation}
\int \alpha(y)\, P_{\eta}(dy) \leq \langle\xi^2\rangle,
\end{equation}
and the equality
\begin{equation}
\int \alpha(y)\, P_{\eta}(dy) = \langle(\xi-\eta)^2\rangle.
\end{equation}
On the other hand, the equivocation, (\ref{eq:equivocation-discretized}),
with fixed outcome $y$ of $\eta$ and given mean $m(y)$ and variance $\alpha(y)$
is maximized if the variable
is Gaussian with variance $\alpha(y)$. Therefore, the mutual information
between $\xi$ and $\eta$ is bounded from below by
\begin{equation}
I(\xi, \eta)\, \geq\, \int \frac{1}{2} \ln \left({{\sigma^2}\over{\alpha(y)}}\right)\, 
       P_{\eta}(dy).
\end{equation}
We can therefore pose a constrained
minimization problem over the auxiliary positive function $\alpha(y)$.
Strictly speaking, $H(\xi,\epsilon)$ is only bounded from below by the
minimization, but, as we will see, the
solution can be realized in terms of variables $\xi$ and $\eta$ with
desired properties.
\begin{eqnarray}
H(\xi,\epsilon) & =& \min_{\alpha(y)}  
           \int \frac{1}{2} \ln \left({{\sigma^2}\over{\alpha(y)}}\right)\,
 P_{\eta}(dy),\label{eq:constrained-minimization} \\
\epsilon^2   &\geq&  \int  \alpha(y)\, P_{\eta}(dy),  \\
 \sigma^2 &\geq&    \int  \alpha(y)\, P_{\eta}(dy).
\end{eqnarray}

Either one or the other of the two constraints will be satisfied at the
minimum in (\ref{eq:constrained-minimization}).
A variation with Lagrange multipliers gives that $\alpha(y)$ must be constant,
either equal to $\epsilon^2$ or to $\sigma^2$.
We therefore have
\begin{eqnarray}
H(\xi,\epsilon)  = \max \,\left[\ln \left({{\sigma}\over{\epsilon}}\right), 0\right].
\end{eqnarray}
The derivation also shows clearly that the solution can be
realized as follows: if $\sigma$ is less than $\epsilon$, then
$\xi | \eta$ is Gaussian with mean zero and variance $\sigma^2$, and $\eta$ a
delta function with support at the origin.
If, on the other hand, $\sigma$ is greater than $\epsilon$, then
$\xi | \eta$ is Gaussian with mean $m(y)$ equal to $y$ and variance $\epsilon^2$, and $\eta$
is Gaussian with mean zero and variance $\sigma^2-\epsilon^2$.
By the semigroup property of Gaussian kernels follows in both
cases that $\xi$ is Gaussian with mean zero and variance $\sigma^2$.

The generalization to higher-dimensional case is
as follows. The variable $\xi$ is Gaussian with second moments
$C_{ij}$, with principal components in a diagonal basis
$(\sigma_1^2,\ldots,\sigma_d^2)$.
At a
point $y_1,\ldots, y_d$ in the support of $P_{\eta}$
we consider the variable $\xi | \eta$ with first moments
$m_i(y_1,\ldots,y_d)$ and second moments  $\alpha_{ij}(y_1,\ldots,y_d)$.
We have, in analogy with the one-dimensional case, 
\begin{equation}
\int \alpha_{ii}(y_1,\ldots,y_d)\, P_{\eta}(dy) \leq \langle\xi_i^2\rangle =  \sigma_i^2,
\label{eq:sigma-equations}
\end{equation}
and 
\begin{equation}
\int \sum_i \alpha_{ii}(y_1,\ldots,y_d)\, P_{\eta}(dy) =  
            \langle(\xi-\eta)^2\rangle \leq  \epsilon^2.
\label{eq:epsilon-equation}
\end{equation}
The mutual information between $\xi$ and $\eta$ is bounded by
\begin{equation}
I(\xi, \eta) \, \geq\, \int \frac{1}{2}\left(
 \ln \det C_{ij} - \ln \det \alpha_{ij}\right)\, P_{\eta}(dy)
\label{eq:mutual-information-Gaussian-manyD}
\end{equation}
Some of the inequalities (\ref{eq:sigma-equations}) may hold as
equalities in the solution, some not. It does however follow from
the diagonal structure of  (\ref{eq:sigma-equations}) and  (\ref{eq:epsilon-equation})
and the relation 
\begin{equation}
\label{eq:neweqapp}
 \delta\ln\det \alpha = \hbox{Tr}\ \frac{\delta \alpha}{\alpha},
\end{equation}
that when (\ref{eq:mutual-information-Gaussian-manyD}) is minimized,
$\alpha_{ij}(y)$ must be constant in $y$ and diagonal in $i$ and $j$.
Let the diagonal elements of $\alpha$ be $D_i^2$. We then have the
following simpler discrete constrained
minimization problem:
\begin{eqnarray}
H(\xi,\epsilon) & =& \min  \sum_i  \ln \left({{\sigma_i}\over{D_i}}\right),  \\
\epsilon^2   &\geq&  \sum_i D_i^2 \label{eq:epsilon-constraint}, \\
 \sigma_i^2 &\geq&    D_i^2.
\label{eq:sigma-constraints}
\end{eqnarray}
The minimum can be found by starting with all $D_i$'s very small
and increasing them proportionally to the gradient, that is,
at the same rate.
When one of the $D_i$'s 
hits the constraint (\ref{eq:sigma-constraints})
we keep it constant from thereon and increase the others.
The constraint  (\ref{eq:epsilon-constraint}) will eventually
be fulfilled when the $D_i$'s that still change are equal to $\theta$, and there we stop.
The solution is thus implicitly given in terms of $\theta$, which can be interpreted
as a cut-off threshold for  modes where normal fluctuations are small:
\begin{equation}
H(\xi,\epsilon)  = \sum_i \max\, \left[\ln\left({{\sigma_i}\over{\theta}}\right), 0\right],
     \qquad 
    \epsilon^2   =  \sum_i \min\,\left[\sigma_i^2, \theta^2\right]
\label{eq:Kolmogorov-formula}
\end{equation}
The solution can be realized by taking $\eta$ a random variable
with independent components, diagonal in the same basis as $\xi$, and letting
the $i$'th component of $\xi$ only depend on the $i$'th component of $\eta$.
The variables $\eta_i$ and $\xi_i |\eta_i$ are then constructed as in
the one-dimensional case, with the only difference that the parameter $\theta$
substitutes for the error $\epsilon$.

\subsection{Gaussian random fields}
Let us now consider a scalar Gaussian random field in $D$-dimensional space.
A particular example ($D=1$) is Gaussian processes. 
We begin by the approximation that the field is periodic with period $L$
in all directions. Fourier components of Gaussian random fields
are independently distributed Gaussian random variables.
Therefore (\ref{eq:Kolmogorov-formula}) can be rewritten,
using the volume element $\Delta  k$,  equal to $(2\pi/L)^D$:
\begin{eqnarray}
H(\xi,\epsilon)\left(\frac{2\pi}{L}\right)^D & =& \sum_{\bbox{k}} \max\, 
              \left[\ln\left({{\sigma_{\bbox{k}}}\over{\theta}}\right), 0\right]\,
\Delta k, \\
\epsilon^2 \left(\frac{2\pi}{L}\right)^D & =&  \sum_{\bbox{k}} \min\, 
                  \left[\sigma_{\bbox{k}}^2, \theta^2 \right]\, \Delta k.
\end{eqnarray}
In the limit when $L$ tends to infinity the right hand sides turn into integrals,
and the left hand sides to entropy and mean-square distance per unit volume.
Therefore we have
\begin{eqnarray}
h^{\hbox{volume}}(\xi,\epsilon) & =& \left(\frac{1}{2\pi}\right)^D\int  
\max\,\left[ \frac{1}{2}\ln\left({{E(\bbox{k})}\over{\theta^2}}\right), 0\right]\, d^D k,
\label{eq:kolmogorov-entropy}  \\
\epsilon^2 & =& \left(\frac{1}{2\pi}\right)^D \int \min\,
\left[E(\bbox{k}), \theta^2\right]\, d^D k.
\label{eq:kolmogorov-error}
\end{eqnarray}
which, for $D$ equal to one, is the Kolmogorov result for the $\epsilon$-entropy
per unit time of a Gaussian random process.

We note that $\theta$ is still a cut-off to eliminate modes such that the energy
of the fluctuations in these modes is less than $\theta^2$.
In fields with a power-law spectrum, such as turbulence, $\theta$ can simply
be substituted for a wave-number cut-off $K$.

\subsection{Other distributions and other fidelity criteria}
\label{ss:other}

The mean square fidelity criterion is convenient for analytical computations,
but it is not the only one possible.
Shannon lists also the maximum distance, another choice would be the
mean absolute distance.

More interestingly, there may be cases where the most appropriate fidelity
criteria is not simply proportional to the distance in the mean or in
maximum value.
If the signal $\xi$ is a L\'evy-process it will have a non-negligible
probability of changing by a large amount over a short interval of time\cite{Levy}.
In many applications one would then
like an approximating signal $\eta$ to capture well the large
jumps, but one would be less interested in a very precise approximation when
$\xi$ changes comparatively little\cite{BouchaudGeorges}.
A reasonable fidelity criterion would then be that there is only a small probability that the
distance between $\xi$ and $\eta$ is large.

As far as we know no investigations have been performed of the  entropy of such sources
with respect to such fidelity criteria. Of course, for a L\'evy process a mean square
error function is not possible, since the second moment is infinite.
It would be possible to use mean absolute error, as was done implicitly by
Gaspard and Wang in \cite{GaspardWang93}, but the relevance of such computation
would have to be motivated in each case by the application at hand.

\section{Spacetime Processes}
\label{s:spacetime}
In this appendix we want to discuss and compare the finite-size Lyapunov exponent
and the $\epsilon$-entropy for a turbulent flow  and for the shell model.
We assume that the flow and the shell model are both 
Gaussian as in appendix~\ref{s:kolmogorov}, and we assume that the spectrum
follows the Kolmogorov theory, as described in section~\ref{s:multifractal}.
Non Gaussian effects and intermittency correction to the spectrum 
are not taken into account.

At a given error threshold $\delta$, we have seen that the finite-size 
Lyapunov exponent scales as 
$\lambda(\delta) \sim \delta^{-2}$, and this holds as
well for the shell model as for $3D$ turbulence.
Let us now consider the $\epsilon$-entropy, and first give a simple dimensional
estimate. It will then be seen that the expressions (\ref{eq:kolmogorov-entropy})
(\ref{eq:kolmogorov-error}) just reproduce this result.
The dimensional estimate starts from the time scales in the Kolmogorov theory,
$\tau(k)\sim k^{-\frac{2}{3}}$. The distance between two fields that only differ
in wave numbers greater than $k$ is $\delta^2 \sim k^{-\frac{2}{3}}$, that is, it does not 
depend on the dimensionality of space, but only of the form of the spectrum
$E(k)\sim k^{-\frac{5}{3}}$.
On the other hand, the number of degrees of freedom at wave numbers less than or
equal to $k$ is proportional to $k^{D}$.
The system must be observed at a rate $\tau(k)^{-1}$ to capture all the motion
in wave numbers less than $k$. The amount of information per unit time and space
needed to describe the
system up to an error tolerance $\delta$ is thus
$k^D\tau(k)^{-1}$. Translating this into a functional dependence on $\delta$ we have
\begin{equation}
h^{\hbox{space},\hbox{time}}(\delta) \sim \delta^{-2-3D}
\label{eq:Gaspard-Wang-scaling}
\end{equation}
which, for $D=3$, leads to $h^{\hbox{space},\hbox{time}}(\delta)\sim \delta^{-11}$.
As far as we can see there is an inconsistency between this result,
also derived in \cite{GaspardWang93}, and the result stated in \cite{GaspardWang92},
which we believe is a misprint resulting from using an argument that really
applies to the finite-size Lyapunov exponent and not the $\epsilon$-entropy.
The very different scaling laws $h\sim \delta^{-11}$ and $\lambda(\delta)\sim \delta^{-2}$
is also a further motivation why introducing the quantity finite-size Lyapunov exponent.

This straight-forward dimensional analysis can easily be generalized to a generic stationary
Gaussian processes with spectral density (see section~6.2.6 of 
\cite{GaspardWang93}),
\begin{equation}
\Phi({\bf k},\omega) \sim k^{-y} F \left({\omega \over k^{z}} \right),
\label{eq:b.3}
\end{equation}
The function $F$ is supposed to vanish for argument much larger and much smaller
than unity, and its integral is on the order of unity.
The energy spectrum is therefore
\begin{equation}
E(k) \sim k^{D-1} \int \Phi({\bf k},\omega) d\omega \sim k^{z-y+D-1}.
\label{eq:b.4}
\end{equation}

We will here check that the previous dimensional estimates can also be obtained
from the Kolmogorov formulae (\ref{eq:kolmogorov-entropy}) and (\ref{eq:kolmogorov-error}),
that we write in this case as follows:
\begin{eqnarray}
h^{\hbox{space}, \hbox{time}}(\delta) & =& \left(\frac{1}{2\pi}\right)^{D+1}\int  
\max\,\left[ \frac{1}{2}\ln\left({{\Phi({\bf k},\omega)}\over{\theta^2}}\right), 0\right]\, d^D k\, d\omega ,
\label{eq:space-time-entropy}  \\
\delta^2 & =& \left(\frac{1}{2\pi}\right)^{D+1} \int \min\,
\left[\Phi({\bf k},\omega), \theta^2\right]\, d^D k\, d\omega .
\label{eq:space-time-error}
\end{eqnarray}

The first observation is that the integral over $\omega$ in (\ref{eq:space-time-error})
gives a contribution of order $k^{z-y}$ if $k^{-y}$ is less than $\theta^2$,
but otherwise a contribution of order $k^z\theta^2$. 
The error is thus determined by a cut-off wave number $K$ such that
\begin{equation}
\delta^2  \sim  K^{z-y+D}\qquad K^{-y}\sim\theta^2
\label{eq:space-time-error2}
\end{equation}

The second observation is that the integral over $\omega$ in 
(\ref{eq:space-time-entropy}) gives a vanishing contribution is 
$k^{-y}$ is less than $\theta^2$, that is, for wave
numbers larger than the cut-off $K$. For smaller wave numbers the 
integration over $\omega$ gives a contribution of the order of $k^z$, 
up to a logarithmic correction that we don't take into account.
The $\epsilon$-entropy with cut-off wave number $K$ is thus
\begin{equation}
h^{\hbox{space}, \hbox{time}}(\delta)  \sim  K^{z+D}\qquad  K^{-y}\sim \theta^2
\label{eq:space-time-entropy2}
\end{equation}

Combining (\ref{eq:space-time-error}) and (\ref{eq:space-time-entropy}) we find
\begin{equation}
h(\delta) \sim \delta^{2\,(D+z)/(z+D-y)}
\label{eq:b.7}
\end{equation}
which is equation~(6.14) of \cite{GaspardWang93}.
By inserting the values $D=3$, $y=13/3$ and $z=2/3$ of the Kolmogorov theory
we reproduce the result 
(\ref{eq:Gaspard-Wang-scaling}).

\end{document}